# Impact of Ethanol and Methanol on NOx Emissions in Ammonia–Methane Combustion: ReaxFF Simulations and ML-Based Extrapolation

Amirali Shateri, Zhiyin Yang, Jianfei Xie *
School of Engineering, University of Derby, DE22 3AW, UK

**Abstract**

The development of ammonia–methane ($NH_3$-$CH_4$) combustion as a hydrogen-carrier energy source faces major challenges such as significant NOx emissions, hindering its practical implementation. This paper examines how ethanol ($C_2H_6O$) and methanol ($CH_4O$) additives influence formation pathways of NOx using ReaxFF molecular dynamics (MD) simulations at temperatures of 2,000 K and 3,000 K. Ten carefully designed fuel mixtures (C1–C10) were evaluated across 0%, 5%, and 10% alcohol concentrations. The findings show that adding alcohol can effectively suppress NOx production, especially at elevated temperatures. At 3,000 K, 10% ethanol addition and 10% methanol addition reduced the production of NOx by approximately 39.6% and 30.1%, respectively, compared with the base fuel. This suppression is attributed to the charge redistribution and the redirection of nitrogen intermediates through stabilising pathways such as HNO, $HNO_2$, and $N_2O$. Simulation-derived descriptors served as the training data for machine learning (ML) models, including Random Forest Regression (RFR), Support Vector Regression (SVR), Gradient Boosting Regression (GBR), and Fully Connected Neural Networks (FCNN). RFR achieved superior performance compared with other models with an R² of 0.993 and mean absolute error (MAE) of 0.661. The trained ML models successfully predicted NOx emissions for both simulated alcohol ratios (0%, 5% and 12%), and non-simulated alcohol ratios (2%, 7%, and 12%) which demonstrating how hybrid physics-informed ML algorithms can extrapolate complex chemical behaviours, along with prediction errors under 5% for most extrapolated ethanol cases. The results showcase that the ReaxFF informed ML framework successfully serves as a basis for designing cleaner fuels and is capable of establishing a reliable structure for future predictive models in combustion chemistry.

**Keywords:** NOx emission; Ammonia–methane combustion; Alcohol additives; Machine learning; Reactive force field; molecular dynamics.

## 1. Introduction

Growing worldwide demand for sustainable energy sources promotes the use of fuels with low carbon emissions and carbon-neutral options. Ammonia ($NH_3$) represents a feasible option for hydrogen transportation and fuel application since it burns carbon-neutrally and can integrate with existing infrastructure [1-2]. The slow reaction kinetics combined with low laminar burning velocity and high ignition temperature make $NH_3$ difficult to use as a standalone fuel. Blending ammonia with hydrocarbon fuels like methane ($CH_4$) has been recommended by researchers to overcome combustion challenges and improve burning process stability and efficiency [3]. Ammonia-methane fuel blends have emerged as promising candidates for low-emission energy systems, especially in high-temperature combustion applications [3]. These fuel blends are being explored for their potential to reduce the environmental impact of combustion processes by lowering harmful emissions, particularly nitrogen oxides (NOx), which are a byproduct of fuel-bound nitrogen oxidation [4].

---

*Corresponding author: j.xie@derby.ac.uk (J.X)

Recent research has revealed how $NH_3$ and $CH_4$ interact synergistically in mixed fuel systems and explored different NOx reduction techniques. Liao et al. [5] demonstrated that $CH_4$ significantly shortens $NH_3$ ignition delay because it accelerates $NH_3$ oxidation through enhanced radical generation under different equivalence ratios and pressure conditions. Kohansal et al. [6] blended ammonia with methane to reduce the laminar burning velocity, but optimal conditions at high temperatures allowed these blends to achieve performance levels comparable to standard hydrocarbon flames with considerable effects on NOx formation process. Studies on the combustion dynamics of $NH_3/CH_4$ mixes are conducted in conjunction with focuses on enhanced burner design. Singh et al. [7] used both computational models and experimental testing in a self-recuperative burner to show how different ammonia levels influence the behaviour of OH, HNO and HCO radicals, which in turn alters flame stability and emission levels of pollutants. Kim et al. [8] showed the use of non-thermal plasma to promote the streamer intensity and enhance the flame propagation in $NH_3/CH_4$-air mixtures, resulting in reduced emissions and better lean-burn performance. In addition, the autoignition characteristics have been investigated for high-temperature behaviour. Chu et al. [9] conducted a study on the ignition delay time of $NH_3/CH_4$ mixtures at temperatures ranging from 1,200 to 2,200 K, indicating consistent ignition performance at specific pressures with $NH_3$-dominant mixtures, thereby suggesting enhanced flame control capabilities for advanced combustion systems. Furthermore, Zhang et al. [10] revealed the unique chain-termination pathways of $NH_3$ in comparison to other fuels under various pressure conditions, contrasting with the high-pressure chemical behaviour of $CH_4$. Han et al. [11] developed empirical correlations between the diluent fraction and laminar burning velocity (SL) for $NH_3$–$CH_4$–$N_2$ mixtures based on the flame structure analysis.

Several studies have shown that incorporating alcohol-based additives into fuel mixtures effectively reduces pollutant emissions, including hydrocarbons (HC), carbon monoxide (CO), and nitrogen oxides (NOx). Usman et al. [12] demonstrated that petrol blends containing ethanol and methanol achieved up to 13.2% reductions in HC emissions with E10 (10% ethanol in 90% of gasoline) blends. When traditional fuels are blended with alcohol fuels, it results in improved brake thermal efficiency (BTE). Pure petrol experienced a 2.35% BTE gain with E10 and a 1.53% with 5% ethanol and 5% methanol (E5M5) blends. Alcohol fuels achieve superior combustion efficiency because of their oxygen content together with high octane ratings. Yu et al. [13] investigated how the methanol affected the kerogen pyrolysis using ReaxFF molecular dynamics (MD) simulations. The study showed that methanol integration changed kerogen's thermal decomposition patterns and changed the creation of essential pyrolysis products. The collective findings from these studies demonstrate how alcohol additives can influence the hydrocarbon pyrolysis process and suggest encouraging applications for fuel optimization in energy solutions [14]. Incorporating alcohol additives into ammonia-methane blends offers enhanced combustion properties.

While alcohol additives such as ethanol and methanol have been promising to improve combustion characteristics and reduce emissions of pollutants, they are highly sensitive to fuel composition, reaction kinetics, and operating conditions in the formation of nitrogen oxides. Because of the complicated nature of the interactions involved, traditional modelling approaches are not sufficient in determining the non-linear relationship present in the NOx formation in ammonia–methane–alcohol systems. This has led to a greater need for tools that can handle correlations between multiple variables and make accurate predictions in a wide range of cases. Machine learning (ML) is a promising solution in this context, facilitating data-driven investigation of combustion behaviour and informing analysis derived from molecular-scale simulations [14]. With the increasing complexity of combustion systems and the need for more accurate emission prediction, ML has become a popular choice as supplementary modelling approach in combustion research. Models such as Artificial Neural Networks (ANN) have been used by researchers to accurately predict NOx emissions, enabling real-time control of combustion parameters and effective mitigation of emissions [15]. In addition, the combination of optimisation algorithms (e.g., genetic algorithms (GA)) with ML platforms has been shown to improve to improve operating conditions, such as temperature, pressure, and fuel–air ratio, leading to more stable and efficient combustion. By combining ANN and GA, it is possible to reduce NOx emissions through optimising excess oxygen concentration and combustion temperature [16]. Similarly, use of ensemble models, such as Random Forest and Extreme Gradient Boosting, has gained popularity due to their robust capacity, with accuracy rates typically being over 90% in the case of gas turbines [17]. The best use of these models is in modelling non-linear and multi-dimensional relations between combustion parameters, which allows generalization over a wide range of fuel types and operating conditions [18-20]. Under the employment of feature optimization methods, their performance can be improved by reducing variable selection and conserving computational capacity. This is particularly important for real-time emission monitoring systems. As combustion systems become more intricate, there is a growing interest in hybrid approaches that combine ML algorithms with physical models, ensuring that predictions are always grounded in chemical and thermodynamic reality. Furthermore, probabilistic modelling approaches—most notably Gaussian Process Regression (GPR)—have demonstrated strong capabilities in quantifying predictive uncertainty and improving model reliability in complex flow dynamics, particularly in cold flow regimes of combustion engines [21].

The present study investigates the effects of alcohol additive substances, ethanol ($C_2H_6O$) and methanol ($CH_4O$), on the formation of NOx in ammonia–methane combustion systems using ReaxFF MD simulations. Simulations were performed at two temperatures (2,000 K and 3,000 K) for three alcohol concentration ratios (0%, 5%, and 10%) to contrast the chemical and thermal impact of additives on NOx emissions. To extend predictive capabilities beyond the limited scope of molecular simulations, an ML platform was used to predict NOx formation for off-design alcohol ratios of 2%, 7%, and 12% that are beyond the original training zone. Unlike most previous ML studies, which focused primarily on interpolation within known datasets or used coarse-grained experimental data as inputs, the present work uniquely combines atomistic simulation-derived descriptors with data-driven learning to enable extrapolative predictions. This is particularly significant in the context of alcohol-blended ammonia fuels, where few studies have rarely addressed how ML models perform in extrapolation regimes, especially for non-simulated fuel compositions. Moreover, while earlier works typically use simplified or global combustion parameters, our study leverages detailed atomic-scale descriptors, such as bond energy distributions, charge dynamics, and species evolution, to inform and train ML models. Different ML models were trained and compared, including Random Forest Regression (RFR), Gradient Boosting Regression (GBR), Support Vector Regression (SVR), and Fully Connected Neural Networks (FCNN), which were experimented with different architectures and validation approaches to identify the most suitable model for extrapolative combustion analysis. Notably, the study recommends a hybrid chemical-guided data-driven approach, whereby interpolated inputs from MD simulations are utilized to generate chemically consistent synthetic datasets. These are then used in NOx prediction by ML without any additional high-cost MD simulations. The proposed hybrid framework is a novel, computationally efficient route to extending combustion behaviour beyond simulated conditions, promoting a reliable modelling strategy that combines atomistic scale information with scalable prediction ability for designing clean fuels.

## 2. Methodology

### 2.1. Reactive force field molecular dynamics (ReaxFF MD)

Reactive force field (ReaxFF) molecular simulations provide a robust framework for modelling chemical reactions at the atomic scale by dynamically accounting for bond formation and breaking. In contrast to traditional force fields based on fixed connectivity, ReaxFF works using bond-ordering and allows for chemical interactions to evolve in real time within highly reactive materials [22-23]. This modelling technique bridges the gap between quantum mechanical computations, which are precise but computationally difficult, and classical molecular dynamics (MD) simulations, which are effective but non-reactive. ReaxFF enables the modelling of extensive systems under extreme pressures and temperatures with a high level of precision, facilitating the physical capture of complex chemical reactions. It has been widely applied in combustion studies, offering valuable insights into reaction mechanisms, intermediate species formation, and product distributions [24]. The ReaxFF can be expresses as a function of the bond order, as described in Eq. (1). The total potential energy in ReaxFF is represented as the sum of various energy contributions, including bond energy, penalties for over- and under-coordination, lone-pair stabilization, valence and torsional angle energies, and non-bonded interactions such as Coulombic and van der Waals forces [25-26]. This formulation allows ReaxFF to accurately capture the complex interplay of bonded and non-bonded interactions in reactive systems. By parameterizing these energy terms based on quantum mechanical calculations, ReaxFF can simulate chemical processes in large systems with both computational efficiency and chemical accuracy.

$$E_{system} = E_{bond} + E_{over} + E_{under} + E_{lp} + E_{angle} + E_{tors} + E_{vdW} + E_{Coul} \qquad (1)$$

### 2.1.1. Validation of ReaxFF

Several reactive force fields are available for the study of C/H/O/N systems, but selecting one that is both accurate and reactive for modelling $NH_3/CH_4$ combustion is critical. For this study, the ReaxFF force fields developed by Kulkarni et al. [27] and Zhang et al. [28] were considered, both of which have been previously validated for ammonia–hydrocarbon systems in the literature [29-30]. To assess the reliability of the selected force field in reproducing reaction energetics, calculated bond dissociation energy (BDE) values were compared with experimental and computational results. The chemical reactions governing $CH_4$ and $NH_3$ combustion involve the primary oxidation pathways of these fuels. Methane ($CH_4$) reacts with oxygen ($O_2$) to produce carbon dioxide ($CO_2$) and water ($H_2O$) as per the reaction:

$$CH_4 + 2O_2 \rightarrow CO_2 + 2H_2O \qquad (2)$$

Simultaneously, ammonia ($NH_3$) undergoes combustion in the presence of oxygen to form nitrogen gas ($N_2$) and water:

$$4NH_3 + 3O_2 \rightarrow 2N_2 + 6H_2O \tag{3}$$

In this study, a 1:1 ratio of $CH_4$ to $NH_3$ is considered, combining these two reactions into a single overall reaction:

$$4CH_4 + 4NH_3 + 11O_2 \rightarrow 4CO_2 + 14H_2O + 2N_2 \tag{4}$$

The reaction pathways for $CH_4$, $NH_3$, and $O_2$ explain the formation of the primary combustion products: $CO_2$, $H_2O$, and $N_2$. The presence of intermediates such as CO, NH, and OH also provides critical insights into the reaction dynamics and the influence of stoichiometric ratios ($\lambda$) on the combustion process.

$$CH_44 \rightarrow CH_3 \rightarrow CH_2 \rightarrow CH \rightarrow C \rightarrow CO \rightarrow CO_2 \tag{5}$$
$$NH_3 \rightarrow NH_2 \rightarrow NH \rightarrow N \rightarrow N_22 \tag{6}$$
$$O_2 \rightarrow O \rightarrow OH \rightarrow H_2O \tag{7}$$

These intermediates are indicative of reaction progress and combustion efficiency under different stoichiometric conditions.

BDE is an important parameter for assessing the reliability of the chosen ReaxFF force field in simulating the reactivity and stability of $CH_4$ and $NH_3$. It is the accurate calculation of the bond dissociation energy and bond order (*BO*) that determines the stability and reactivity of molecules in combustion systems. BDE provides a quantitative estimate of the amount of energy required to deform a given chemical bond, and this is one of the most important parameters for calculating reaction energetics. The ReaxFF framework calculates bond energy (*E*) as a function of the bond order (*BO*), which represents the degree of bonding between two atoms [26, 31].

$$E = -D_e \cdot BO \cdot \exp\left(p_{be2} \cdot \left(1 - BO^{p_{be1}}\right)\right) \tag{8}$$

$$BO = \exp\left(p_{bo1}\left(1 - \left(\frac{r}{r_0}\right)^{p_{bo2}}\right)\right) \tag{9}$$

The parameters $pbo1$ and $pbo2$ control how bond order decays with bond stretching or compression relative to a reference bond length $r_0$. The parameters $pbe1$ and $pbe2$ control how bond energy depends on bond order. *De* represents the dissociation energy for the bond. This equation ensures that energy scales with bond order, reflecting the bond's contribution to the system's total energy. In this study, the bond dissociation energy (BDE) values for $CH_4$ and $NH_3$ were calculated using the ReaxFF 2009 force field developed by Zhang et al. [28] and were compared against both experimental data [32–33] and simulation results from Xu et al. [29], which used the ReaxFF 2012 force field by Kulkarni et al. [27]. For $CH_4$, the BDE obtained in the present study was 104.35 kcal/mol, compared to 105.0 kcal/mol in Xu et al. [29] and 102.6 kcal/mol experimentally. This results in an absolute error of 0.65 kcal/mol versus Xu et al. and 1.75 kcal/mol versus the experimental reference. For $NH_3$, the present study yielded a BDE of 106.09 kcal/mol, while Xu et al. reported 107.4 kcal/mol and the experimental value was 103.6 kcal/mol. This corresponds to absolute errors of 1.31 kcal/mol and 2.49 kcal/mol, respectively. These results indicate that the ReaxFF 2009 force field produces BDE values in close agreement with both experimental benchmarks and previously published simulations, supporting its suitability for accurately capturing bond-breaking energetics in ammonia–methane combustion systems. The near equivalence of $CH_4$ and $NH_3$ BDE values also aligns with their similar molecular stabilities, as reflected in their comparable consumption rates during combustion. This supports the applicability of the ReaxFF. 2009 (Zhang et al. [28]) ReaxFF force field for exploring complex reaction mechanisms and product formations in $CH_4/NH_3$ systems under varying conditions.

2.1.2. Cases set-up

To investigate the influence of alcohol additives on ammonia–methane combustion, ten computational cases (C1–C10) were configured. Each case employed an equivalence ratio of $\lambda = 0.7$, a moderately fuel-rich condition typical of practical combustors where perfect stoichiometry ($\lambda = 1$) is rarely achieved. Operating under fuel-rich conditions promotes the build-up of reactive intermediates and enables detailed examination of NOx formation pathways which is the primary objective of this study. For physical consistency, the mass density was fixed at 0.34 g/cm³ in every simulation. Although the total number of molecular remained constant at 800, the cubic domain length was adjusted case-by-case to satisfy the density constraint, ensuring that any changes in system behaviour originate solely from variations in fuel composition or temperature. The baseline mixture comprised 422 fuel molecules—methane ($CH_4$) and ammonia ($NH_3$) in a 1:1 ratio—together with 378 oxygen molecules. Alcohol co-fuels, ethanol ($C_2H_6O$) or methanol ($CH_4O$), were introduced by substituting 5 % or 10 % of the original $CH_4$-$NH_3$ base fuel while preserving the overall molecule count. Distinct C–H bond energies and oxygen contents in selected alcohols are expected to affect their radical pools and chain-branching behaviour. Each composition was evaluated at 2000 K and 3000 K to capture temperature-dependent kinetics.

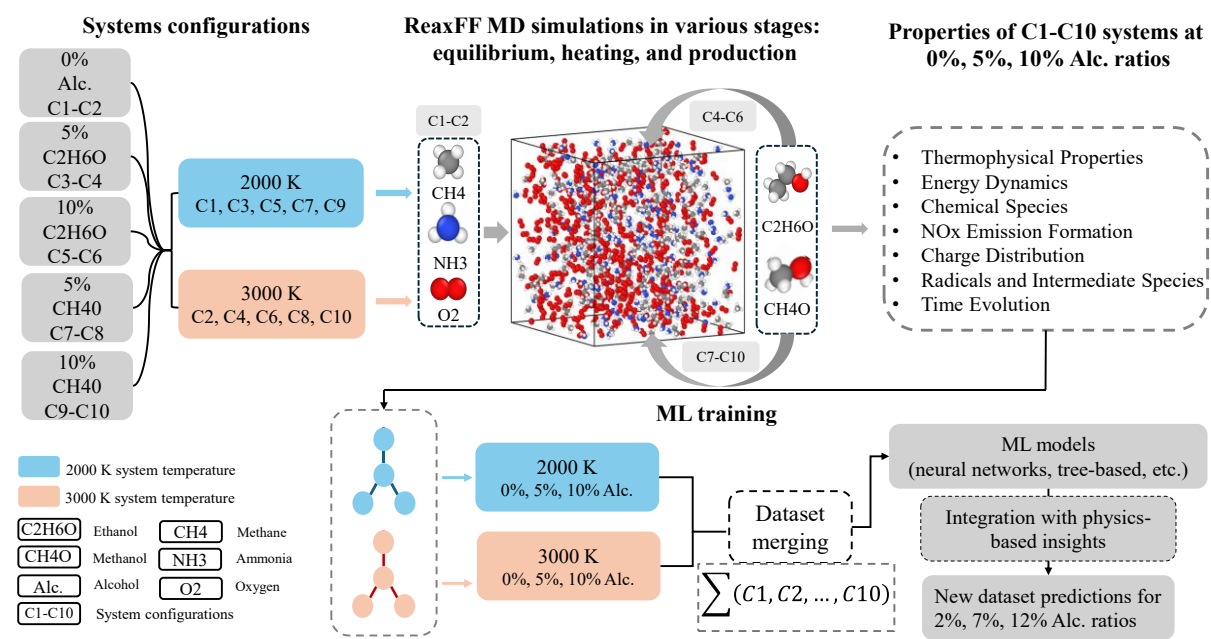

**Fig. 1.** Schematic representation of case configurations, fuel compositions, and simulation workflow for ammonia–methane combustion systems with alcohol additives. The diagram outlines the formulation of ten cases (C1–C10) at 2000 K and 3000 K with varying alcohol ratios (0%, 5%, 10%), molecular substitutions, and subsequent integration into a merged dataset for machine learning prediction.

Tables S1 and S2 in SI material list the full set of simulation parameters, including density, temperature, molecular composition, and corresponding cube size for every case. These data establish a consistent framework for quantifying how incremental alcohol addition modulates intermediate-species production and, ultimately, NOx emissions under fuel-rich combustion. A visual summary of the case configurations is presented in Fig. 1, which outlines the temperature–composition matrix, the fuel replacement scheme, and the corresponding simulation domains. The diagram also highlights how the individual systems were integrated into a single dataset, later used for machine learning predictions targeting off-design alcohol ratios.

2.1.3. Simulation set-up

MD simulations were conducted using LAMMPS (Large-scale Atomic/Molecular Massively Parallel Simulator) [34], a widely used molecular dynamics platform that supports both reactive and non-reactive force fields, including ReaxFF for chemical reaction modelling. Simulations were executed on a multi-core compute node to achieve sufficient statistical sampling within a feasible wall-clock time. Initial configurations were assembled in PACKMOL [35], ensuring uniform spatial distribution, randomised molecular orientation, and avoiding initial molecular overlaps. Reactive interactions were described with the ReaxFF potential; the C/H/N/O parameterisation of Zhang et al. [28] was chosen because it reliably reproduces ammonia–methane chemistry and the associated NOx forming intermediates.

Charge equilibration was carried out using the qeq/reax method at every timestep, and a bond order cutoff of 0.3 was used to detect significant chemical interactions [29, 30]. The simulation workflow comprised three key stages: equilibration, heating, and production. After geometrical relaxation, the system was equilibrated for 100ps in the NVT ensemble at 300 K with a 0.1fs timestep. The temperature was then raised linearly from 300 K to the target value—either 2000 K or 3000 K—at 20 K.ps$^{-1}$; during this ramp, the timestep was reduced to 0.05 fs to accommodate increased atomic velocities. The main production run was performed for 500 ps at the target temperature, employing a 0.01 fs step length to resolve high-frequency vibrational modes and reactive collisions accurately [49, 50]. All dynamic stages retained the NVT ensemble to maintain constant temperature-volume conditions [29, 30, 48]. Trajectory data were analysed with the REAXC package embedded in LAMMPS to monitor reaction pathways and bond-network evolution. Post-processing was carried out with ChemTraYzer [36] supplemented by in-house Python scripts, enabling automatic species identification, pathway analysis, and quantitative evaluation of NOx formation. Atomistic trajectories were visualised with OVITO [37] to facilitate qualitative inspection of structural changes.

2.2. Machine learning framework

2.2.1. Data structure and preprocessing

Fig. 2 shows the data processing pipeline that links MD simulations to ML models applied in this paper. ReaxFF-based MD simulations created the baseline dataset from simulations performed at temperatures 2,000 K and 3,000 K for 10 chemical cases, which are labelled C1 to C10. Each case simulated a 500ps reactive trajectory and produced a series of thermochemical and molecular descriptors, which were designed to characterise the evolving behaviour of ammonia–methane–alcohol mixtures. The resulting dataset includes 26 input features ($X$), encompassing key thermophysical observables such as temperature, pressure, and density, as well as energy terms including total, kinetic, and potential energies. In addition, features derived from the ReaxFF were included to represent mechanistic quantities of the direct chemical relevance. These consist of individual energy contributions: v_ea (atomic), v_eb (bond), v_elp (lone pair), v_ev (valence angle), v_epen (penalty), v_ecoa (conjugation), v_ehb (hydrogen bond), v_et (torsion), v_eco (conjugated), v_ew (van der Waals), v_ep (Coulomb), and v_eqeq (charge equilibration). Element-specific charge distribution properties were also computed: v_qC, v_qH, v_qN, and v_qO, representing the mean partial charge for each atom type, which are essential for capturing electron redistribution during combustion. Chemical identity was represented by the mass fraction of alcohol feature, which distinguishes 0%, 5%, and 10% mixed fuels, and binary indicators for ethanol and methanol, to account for molecular structure effects. These enable the models to learn not only compositional trends but also the functional impact of hydroxyl content and C–H/O–H bond energetics on various alcohol additives. A rigorous data cleaning step was employed prior to training the models. This included the detection and removal of anomalous entries owing to rare bond breakage events, unphysical charge spikes, and species mislabelling during dynamic equilibration. Outliers were eliminated using interquartile-based cutoff values, and no missing values were encountered, as molecular dynamics simulations deterministically generate complete trajectories with time-resolved data for all atoms and properties. Preprocessing was followed by 80:20 (e.g., 80% of data used for training and 20% for validation) stratified train-test splitting in a manner that ensured proportional representation of all the compositions and temperatures within the training and testing sets [14, 38].

This balance was critical to support downstream generalisation, especially in light of the hybrid extrapolation framework, which is introduced later. As illustrated in Fig. 2, the data served as input to four different ML architectures, e.g., Random Forest Regression (RFR), Gradient Boosting Regression (GBR), Support Vector Regression (SVR), and a Fully Connected Neural Network (FCNN). The NOx output ($Y$) was predicted both within the training range (0%, 5%, 10%) and extrapolated beyond (e.g., 2%, 7%, 12%) using a post-processing formulation grounded in chemical trends. To determine nitrogen oxide emissions from each system, the mole fractions of individual $NO_x$-related species (NO, $NO_2$, $N_2O$, and $HNO_3$) were extracted from the LAMMPS ReaxFF simulation outputs using the fix reax/c/species command. These species were tracked over time, and their aggregate concentrations were used to estimate the total $NO_x$ emission in parts per million (ppm). The total $NO_x$ was computed using the following formulation:

$$NO_X = \left( \frac{\sum_{i \in NO_X \, species} n_i}{N_{total}} \right) \times 10^6 \qquad (10)$$

Where $n_i$ represents the number of molecules of each $NO_x$ species (i.e., NO, $NO_2$, $N_2O$, $HNO_3$) identified in the system, $N_{total}$ is the total number of molecules in the simulation box. This equation follows standard post-processing protocols for ReaxFF molecular dynamics simulations used in combustion and pollutant analysis [23-24]. The multiplication by 106 converts the molar ratio into a ppm-scale value for better interpretability and alignment with experimental conventions.

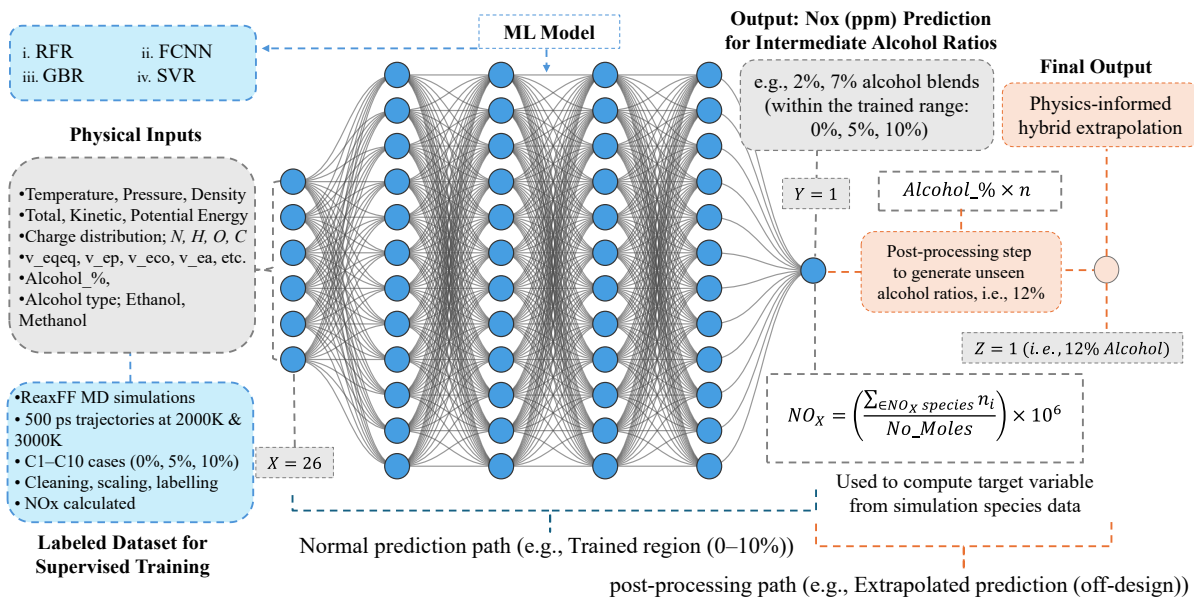

**Fig. 2.** Machine learning (ML) workflow integrating chemical-guided extrapolation for NOx prediction from ReaxFF-derived combustion data. NOx was calculated based on the normalized counts of NOx-related species (i.e., NO, $NO_2$, $N_2O$), as described in Eq. (10)

2.2.2. ML algorithms

Four ML algorithms were chosen in this work to model the intricate interaction between combustion descriptors and NOx emissions: RFR, GBR, FCNN, and SVR. Their capacity to capture non-linear dependencies, resilience against overfitting and proven success in thermochemical prediction tasks guided selection of these models. All the algorithms have different strengths that are well-suited to the nature of combustion-related data. RFR, as an ensemble decision tree, performs exceptionally well with high-dimensional data and produces variable importance information, which is critical in understanding the influence of physical parameters. GBR augments the same by incorporating gradient-based optimization in the construction of trees and thereby is well suited to pick up weak interactions in noisy or structured data. FCNNs, on the other hand, are capable of learning intricate, continuous functions due to their hierarchical representation power, particularly useful in tracking multi-variable chemistries' dynamics. SVR becomes a standard regression benchmark because it excels with non-linear data by using the kernel-based similarity functions to capture non-linearity. Four distinct versions of each algorithm were created to explore the effect of design choices on their performance by adjusting internal parameters. To evaluate the robustness and generalizability of tree-based algorithms (i. e., RFR, GBR), a systematic approach was taken to adjust both tree count and depth parameters. Neural networks (NN) configurations included various hidden layer counts and neuron distributions ranging from basic one or two layered networks to deep structures with up to five hidden layers. SVR configurations were altered by experimenting with different kernel functions (linear, polynomial, or radial basis) and regularisation parameters.

Table 1 summarises the architectural configurations used across all models. This systematic variation allows a comprehensive evaluation of how model complexity and hyperparameter tuning influence the prediction accuracy in a reactive system. By comparing models not only across algorithm types but also across architectures, the present study aims to determine which models are most effective at extracting meaningful chemical reactions from ReaxFF-generated data and generalising to unobserved alcohol ratios in a chemical-guided extrapolation setting.

**Table 1** Summary of hyperparameter variations and architectures for Random Forest Regression (RFR), Gradient Boosting Regression (GBR), Fully Connected Neural Network (FCNN), and Support Vector Regression (SVR) models in predicting NOx emissions.

### FCNN Model Configurations

| Model | Search Method | Models Compared | NN Shape | Activation | Dropout | Batch Size | Training Steps | CV Strategy | Folds | Scoring |
|---|---|---|---|---|---|---|---|---|---|---|
| FCNN 1 | - | - | [100, 100] | ReLU | 0.05 | 64 | 1000 | K-Fold | 5 | RMSE |
| FCNN 2 | Randomised | 50 | [128, 128, 128, 128, 128] | Swish | 0.05 | 128 | 1000 | K-Fold | 5 | RMSE |
| FCNN 3 | Randomised | 50 | [64, 64, 64] | ReLU | 0.1 | 32 | 1000 | K-Fold | 5 | RMSE |
| FCNN 4 | Randomised | 50 | [256, 256, 256, 256, 256] | Swish | 0.05 | 256 | 1500 | K-Fold | 5 | RMSE |

### RFR Model Configurations

| Model | Search Method | Models Compared | Number of Estimators | Max. Depth | Min. Split Samples | Min. Leaf Samples | Grid Estimators | Grid Max. Depth | Grid Min. Split | Grid Min. Leaf | CV Strategy | Folds | Scoring |
|---|---|---|---|---|---|---|---|---|---|---|---|---|---|
| RFR 1 | - | - | 100 | None | 2 | 1 | None | None | None | None | K-Fold | 5 | RMSE |
| RFR 2 | Randomised | 20 | 600 | 20 | 2 | 2 | [700,300,400,500,600] | 12,15,18,20 | 2,4,8,16 | 2,4,8,16 | K-Fold | 5 | RMSE |
| RFR 3 | Randomised | 50 | 400 | 20 | 4 | 2 | [200,300,400,500,600] | 12,15,18,20 | 2,4,8,16 | 2,4,8,16 | K-Fold | 5 | RMSE |
| RFR 4 | Randomised | 30 | 300 | 20 | 16 | 2 | [100,200,300,400,500] | 10,20 | 2,4,8,16,32 | 2,4,8,16,32 | K-Fold | 5 | RMSE |

### SVR Model Configurations

| Model | Search Method | Models Compared | Kernel | Regularisation Parameter (C) | Kernel Coefficient (Gamma) | Grid (C) | Grid (Gamma) | CV Strategy | Folds | Scoring |
|---|---|---|---|---|---|---|---|---|---|---|
| SVR 1 | - | - | RBF | 1 | Scale | None | None | K-Fold | 5 | RMSE |
| SVR 2 | Randomised | 10 | RBF | 100 | Scale | [0.1, 1, 10, 100, 1000] | scale, 0.01, 0.1 | K-Fold | 5 | RMSE |
| SVR 3 | Randomised | 20 | RBF | 100 | Scale | [0.001, 0.01, 0.1, 1, 10, 100] | scale, auto, 0.0001, 0.01, 0.1 | K-Fold | 5 | RMSE |
| SVR 4 | Randomised | 20 | Polynomial | 1000 | 0.1 | 1000 | scale, 0.01, 0.1 | K-Fold | 5 | RMSE |

### GBR Model Configurations

| Model | Search Method | Optimised Parameters | Number of Estimators | Learning Rate | Max. Depth | Min. Split Samples | Min. Leaf Samples | Subsample | Grid Max. Depth | Grid Min. Split | Grid Min. Leaf | CV Strategy | Folds | Scoring |
|---|---|---|---|---|---|---|---|---|---|---|---|---|---|---|
| GBR 1 | - | No | 500 | 0.05 | 6 | 2 | 1 | 1 | None | None | None | K-Fold | 5 | RMSE |
| GBR 2 | Grid Search | Yes | 300 | 0.01 | 8 | 8 | 4 | 1 | 3 | 5 | 4 | K-Fold | 5 | RMSE |
| GBR 3 | Grid Search | Yes | 800 | 0.03 | 6 | 4 | 4 | 1 | 10 | 10 | 6 | K-Fold | 5 | RMSE |

| GBR | 4 | Grid Search | Yes | 1000 | 0.05 | 4 | 2 | 2 | 0.8 | 6 | 8 | 4 | K-Fold | 5 | RMSE |

2.2.3. Model evaluation

Performance comparison of different ML models employed mean absolute error (MAE), mean squared error (MSE), Pearson correlation coefficient (PCC), and coefficient of determination ($R^2$) as regression evaluation metrics. MAE calculates the average error magnitude while MSE determines the average error dispersion with MSE being more affected by outliers than MAE [39–40]. PCC and $R^2$ represent the correlation between predicted results and actual data points using their correlation measurement [41–42]. All the models were trained using 80% of the data to keep the remaining 20% as a test set for a fair comparison and generalization [38]. The analysis of the absolute error values and correlation metrics of the training and test sets can be used to identify overfitting and to choose the best model to interpret physical and chemical principles.

2.3. Chemical-guided extrapolation

To extend the predictive capability of different ML models beyond the training range, a chemical-guided extrapolation strategy was employed. This approach aimed to infer NOx emissions for untrained alcohol ratios, specifically 2%, 7%, and 12%, by leveraging both data-driven models and domain-specific chemical insights derived from ReaxFF simulations. As illustrated on the right panel of Fig. 2, the post-processing workflow builds upon the trained model's outputs at known alcohol concentrations (0%, 5%, and 10%). The key idea was to avoid the need for new MD simulations at unseen compositions, and instead to estimate the target variables by interpolating or extrapolating trends observed in the training data. This hybridisation of ML and physical chemistry offers a practical route for off-design predictions without incurring the computational cost of generating new datasets. Specifically, for in-range predictions such as 2% and 7%, a linear interpolation scheme was adopted between the adjacent training points (e.g., 0%–5%, 5%–10%). This approach assumes smooth transitions in NOx behaviour across small alcohol increments, which was consistent with the gradual trends captured by ReaxFF simulations. For out-of-range prediction at 12%, which is a value not seen during model training, a scaling-based extrapolation technique was applied. This was informed by the observed gradient between the 5% and 10% alcohol systems and adjusted proportionally beyond the upper bound. To mathematically implement this extrapolation, a scaling coefficient is introduced and is expressed below:

$$Feature_{12\%} = Feature_{10\%} + \left(\frac{12-10}{10-5}\right) \times (Feature_{10\%} - Feature_{5\%})  \tag{11}$$

This transformation was applied independently to each input feature before feeding it through the trained models, allowing the prediction of NOx emissions for novel alcohol levels. This method is described in Fig. 2 as post-processing step to generate unseen alcohol ratios, and it bridges the gap between chemical-guided trends and data-driven modelling. The underlying rationale for this strategy is based on the smooth, thermochemically governed nature of alcohol substitution in reactive systems. Then, ML models, which were trained on clean, labelled inputs, were able to capture the nonlinear coupling between chemical features and NOx formation, reinforcing the robustness of predictions. By combining these two pillars, e.g., data-driven learning and chemically grounded extrapolation, the hybrid method demonstrated in this study offers a reliable alternative to conventional simulation-heavy routes for combustion design. This novel approach enables fast, cost-effective insights into emission reduction under off-design conditions, showcasing a vital step towards adaptive, intelligent optimisation in combustion chemistry.

## 3. Results and Discussion
3.1. MD simulation results
3.1.1. Charge distribution trends

The influence of alcohol additives on charge redistribution during ammonia–methane combustion was assessed by monitoring the ReaxFF partial charges of nitrogen (N), hydrogen (H), carbon (C), and oxygen (O) atoms. Figs. 3 and 4 present the average charge–time profiles for all mixtures (0%, 5%, and 10% alcohol) at 2,000 K and 3,000 K, respectively. At 2,000 K in Figs. 3(a–d), the nitrogen sites acquire progressively more positive charge in every mixture, the rise being most pronounced for the methanol-containing cases C7 and C9. Enhanced electron withdrawal from nitrogen in these systems points to an alcohol-assisted oxidation route for $NH_3$, which is consistent with the hydroxyl-rich environment created by methanol. Hydrogen, by contrast, maintains an almost constant charge, with only minor increases observed at higher alcohol ratios; this stability suggests a passive role for H atoms during the early radical-initiation period. Carbon behaviour depended strongly on additive type and ratio. The reference mixture C1 (0% alcohol) retained the most negative carbon charge throughout, whereas the richer ethanol and methanol blends (C5 and C9) drifted gradually toward neutrality. The shift is attributed to the greater oxygen content of alcohols, which pulls electron density away from carbon as additional C–O bonds form. A parallel trend emerged for oxygen: in all alcohol-rich mixtures, the O atoms became steadily more negative, reflecting growth of partially reduced species, principally OH groups and nascent NOx intermediates.

At the elevated temperature of 3,000 K in Figs. 4(a–d), the partial charge dynamics evolve significantly. Throughout the run, the net charge on nitrogen declines and in case C6, becomes slightly negative, indicating very rapid ammonia dissociation and prompt conversion toward nitric oxide. Hydrogen partial charge rises in every mixture, most clearly in the alcohol-rich systems, illustrating a pattern that reflects faster hydrogen-abstraction reactions and elevates OH and H radical production. C atoms exhibit a general trend toward increased positivity (see Fig. 4(c)), showing the steepest climb in C10. This behaviour supports the idea that high thermal energy promotes extensive carbon oxidation, especially in methanol-enriched mixtures. Oxygen again becomes more negative and more quickly than that at 2,000 K, confirming that the 3,000 K environment is substantially more reactive.

Collectively, these temperature-dependent charge shifts underscore the delicate balance between the fuel composition and radical chemistry. Methanol, in particular, intensifies electron redistribution, which is a factor that is expected to influence the subsequent NOx-formation pathways discussed in the following section.

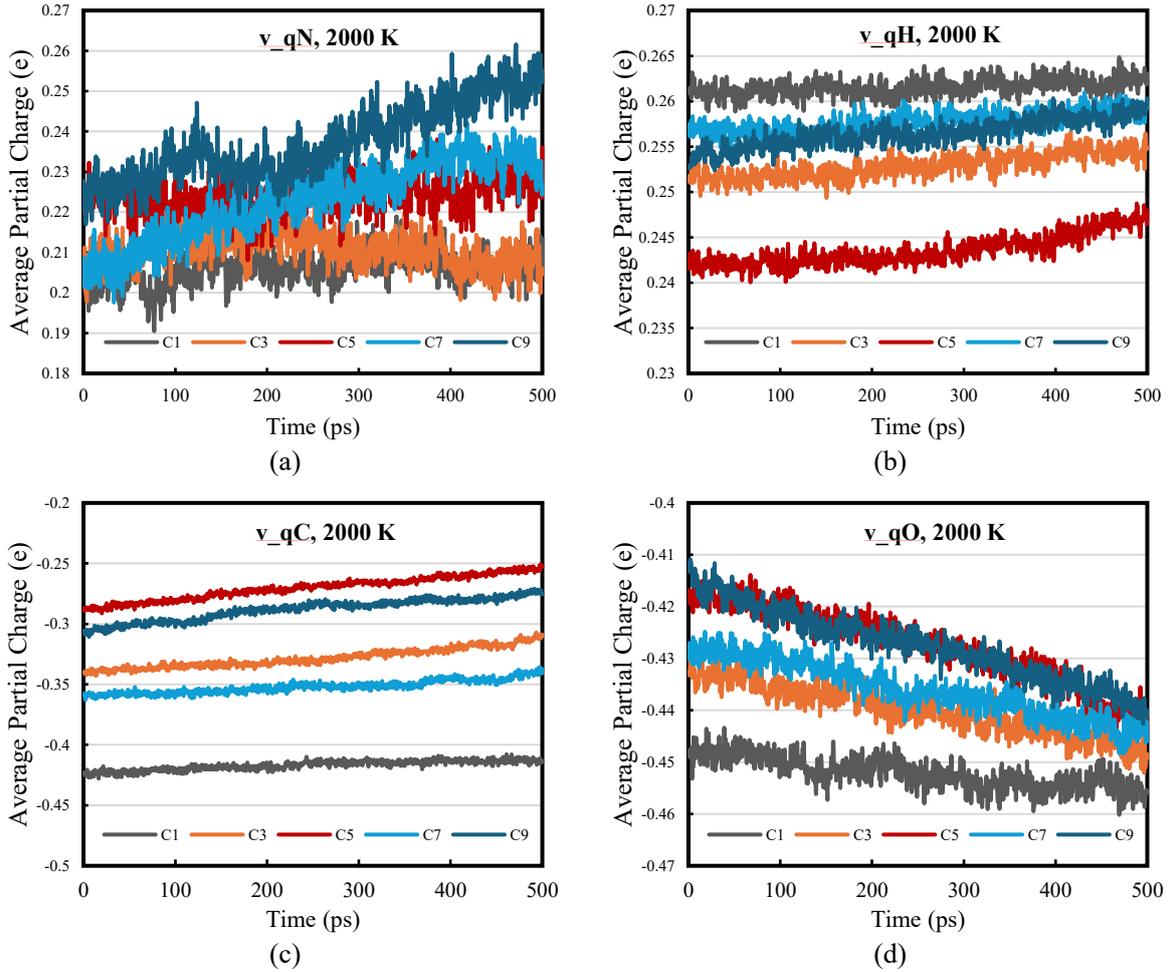

**Fig. 3** Time evolution of average partial charges for nitrogen (N), hydrogen (H), carbon (C), and oxygen (O) atoms at 2,000 K across C1, C3, C5, C7, C9 cases: (a) v_qN; (b) v_qH; (c) v_qC; and (d) v_qO.

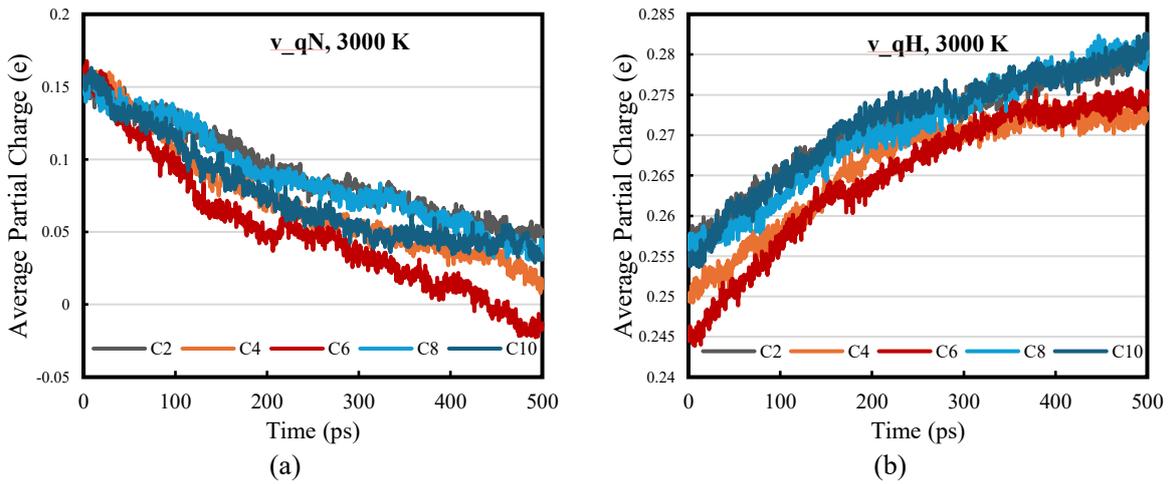

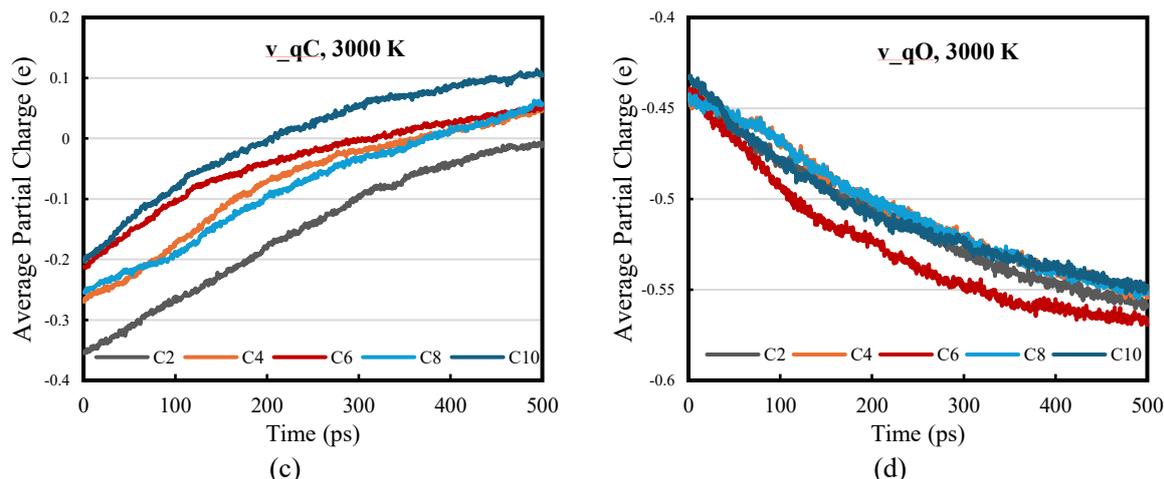

**Fig. 4** Time evolution of average partial charges for nitrogen, hydrogen (H), carbon (C), and oxygen (O) atoms at 3,000 K across C2, C4, C6, C8, C10 cases: (a) v_qN; (b) v_qH; (c) v_qC; and (d) v_qO.

3.1.2. Charge equilibration and NOx emissions

Charge equilibration energy in ReaxFF simulations reveals the electron redistribution during chemical processes. Fig. 8 displays how this parameter changes during combustion at 2,000 K (see Fig. 5(a)) and 3,000 K (see Fig. 5(b)), highlighting the role of alcohol additives in forming dynamic electronic structures.

During the simulation at 2,000 K, mixtures that contain methanol show significantly lower charge equilibration energies when compared with other mixtures. The system demonstrates a more stable electron distribution, which can be primarily attributed to the larger number of electronegative oxygen atoms present in alcohol molecules. The single C atom and pronounced polarity of methanol lead to early polar interactions that accelerate the electron delocalisation and subsequently decrease the excessive charge build-up on reactive intermediates [43]. The base fuel (C1, 0% Alc.) demonstrated the maximum equilibration energy, suggesting a less favourable charge distribution and probably slower radical reactions. The influence of alcohol additives becomes even more pronounced at 3,000 K (see Fig. 5b). At this elevated temperature, ethanol-containing mixtures, particularly C6 (10% ethanol), achieved the maximum equilibration energy among all cases, underscoring their high reactivity and delayed stabilisation of the charge network. Conversely, methanol-rich systems, such as C10, show more moderate equilibration trends, indicating a superior capacity for sustaining intermediate species without excessive charge localisation. The increased thermal energy in these cases amplifies polar bond interactions to accelerate charge redistribution processes, especially in oxygen-rich molecular environments.

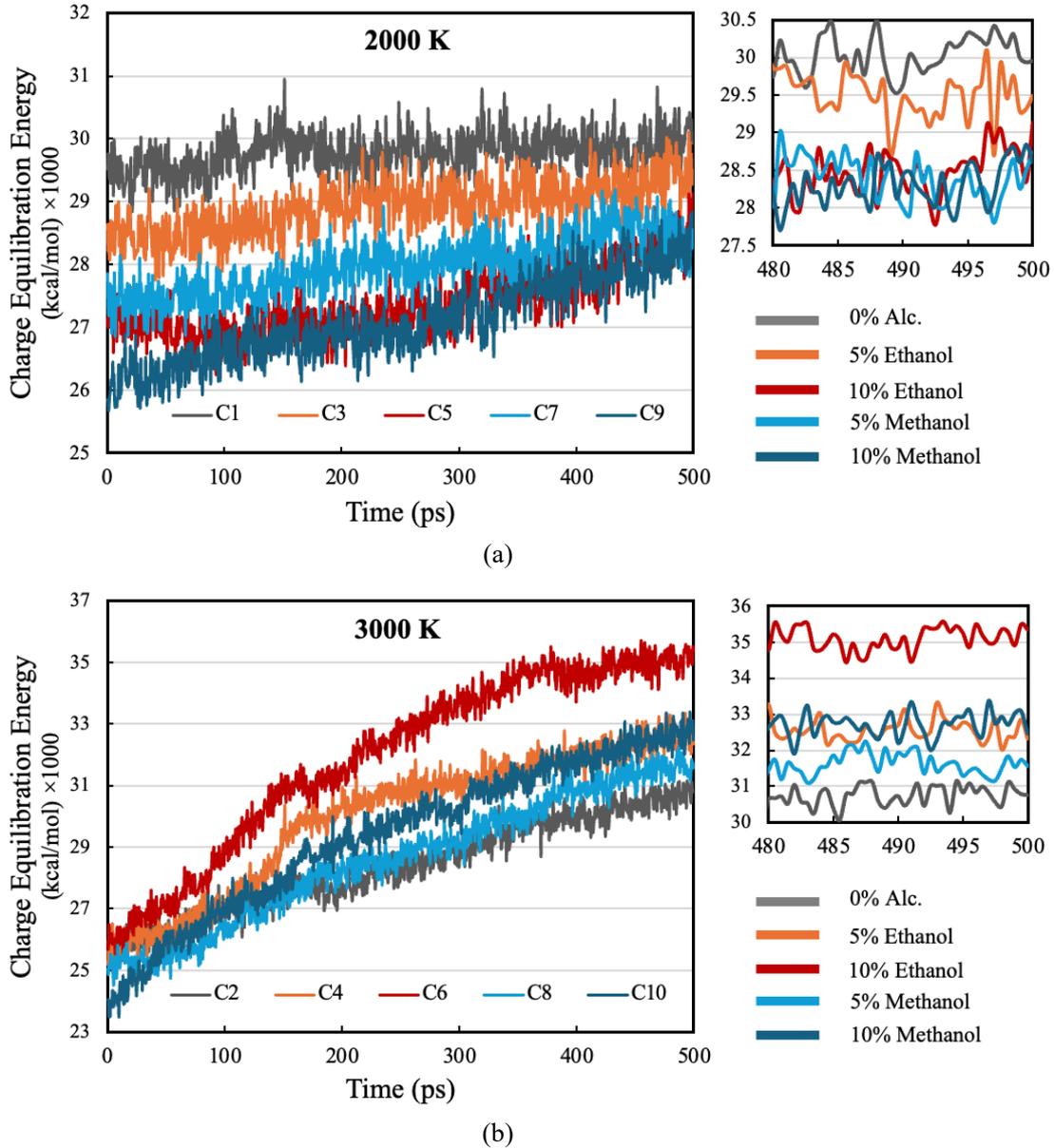

**Fig. 5** Temporal evolution of charge equilibration energy for all fuel blends (0%, 5%, 10% Alc.) at: (a) 2,000 K and (b) 3,000 K. Insets magnify the final 20ps to highlight the equilibrium behaviour.

A direct connection can be observed between charge equilibration dynamics and NOx formation trends, as displayed in Fig. 6. At 2,000 K in Fig. 6(a), NOx emissions are highest in methanol-containing blends, particularly C7 (5%) and C9 (10%), corresponding closely to the steepest charge gradient observed in those mixtures. The connection between electron-rich conditions caused by polar additives and early NO formation demonstrates that rapid $NH_3$ oxidation drives this process [44]. At 500ps, the base fuel C1 produces a NOx level of 90.79ppm; conversely, C5 including 10% ethanol obtains the lowest NOx concentration at 88.31ppm. At 100.76ppm and 110.54ppm respectively, C3 with 5% ethanol and C7 with 5% methanol produce higher NOx emissions; C9 with 10% methanol produces 94.46ppm. Using 10% methanol in fuel results in lower NOx emissions compared with the base fuel although NOx emissions rise with other alcohol additives. At 3,000 K in Fig. 6(b), however, the NOx profiles reverse. The blend with the highest ethanol content demonstrates minimal NOx output, illustrating a change in dominant chemical reaction routes. Here, the faster oxidation kinetics enabled by high temperature and coupled with the stabilising characteristic of ethanol's molecular structure, appear to suppress NO formation. Methanol-containing mixtures, on the other hand, remain more reactive over time, with higher NOx levels sustained at the final 20ps (see inset in Fig. 6(b)). The base fuel (C2) produces 109.95ppm of NOx at 500ps while C6 with 10% ethanol results in the lowest NOx level at 66.46ppm, which is followed by C10 with 10% methanol at 76.84ppm, then C4 containing 5% ethanol at 81.63ppm, and C8 with 5% methanol at 89.99ppm. Methanol promotes NOx formation via enhanced reactivity and rapid oxygen donation, primarily through $NO_2$ and $HNO_3$ intermediates. In addition, alcohol-derived radicals such as OH become more active at high temperatures to interact with nitrogen-containing intermediates, thereby stopping the forming routes of NOx [46-47]. NOx concentration inversions caused by variations in temperatures demonstrate how various alcohol types and concentrations affect this phenomenon. Methanol leads to early NOx production when the temperature is moderate. However, ethanol reduces NOx output at high temperatures even though it contains more energy. At 3,000 K, 10% ethanol reduces NOx emissions by approximately 39.5% compared to the base fuel, while 10% methanol achieves a 30.1% reduction. These findings underscore ethanol's superior role in suppressing NOx at high temperatures. It can be concluded that alcohol additives, particularly ethanol, emerge as promising solution for cleaner combustion by lowering emissions while supporting thermal efficiency improvements. This transition in NOx trends from 2,000 K to 3,000 K reflects a fundamental shift in dominant chemical mechanisms.

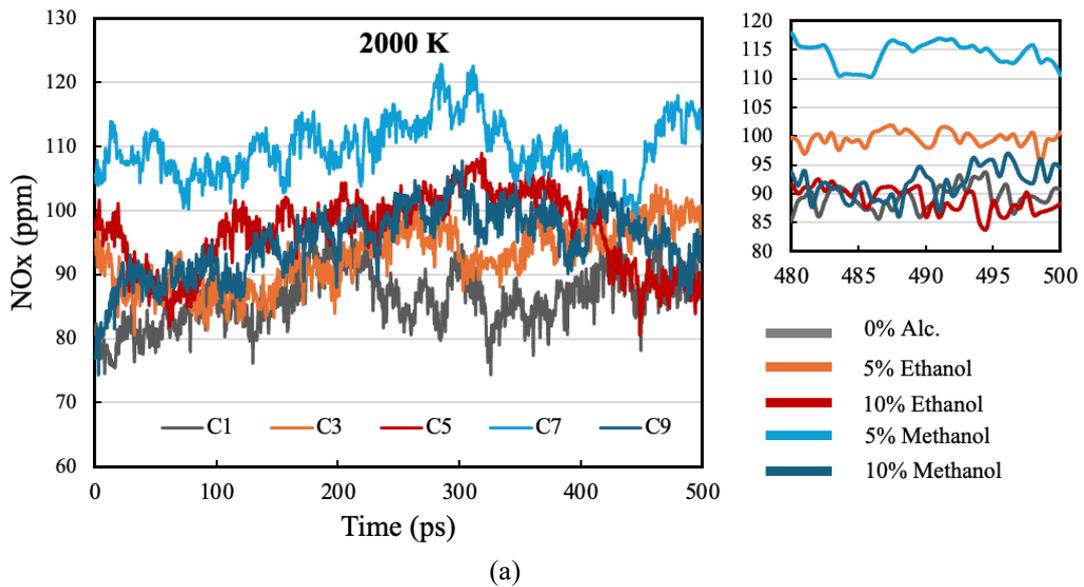

(a)

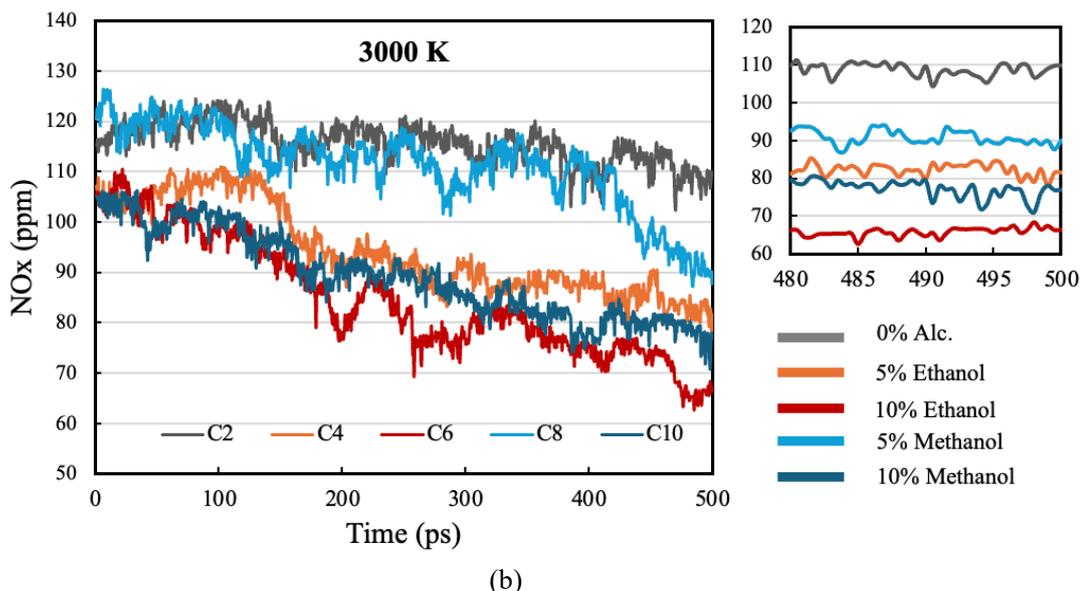

(b)

**Fig. 6.** NOx emissions over time for: (a) 2,000 K and (b) 3,000 K cases. Trends are directly linked to alcohol type (ethanol and methanol) and concentration (0%, 5%, 10% Alc.). Insets show stabilisation levels at the final 20ps of the simulation.

### 3.1.3. Dominant reaction pathway

To elucidate how alcohol additives modulate nitrogen–oxygen reaction networks during combustion, Table 2 compiles the forward and reverse frequencies of the most active $NO_x$-associated reactions across all ten fuel systems (C1–C10). The reactions were identified based on their participation in NO, $NO_2$, HNO, $HNO_2$, and $HNO_3$ interconversion loops, critical mechanisms governing $NO_x$ formation and removal. As observed, the base case (C1, 2000 K) exhibits strong engagement of $NO_2 \rightleftharpoons HNO_3$ and $HNO_3$-related formation pathways, reflecting the dominance of higher-order nitrogen oxides under lower-temperature oxidative conditions. At 3000 K (C2), the reaction network shifts toward simpler cycles such as $HNO_2 \rightleftharpoons NO$ and $NO \rightleftharpoons HNO$, indicating increased fragmentation of $NO_x$ species and reduced stability of nitric acid derivatives at elevated temperatures. The addition of alcohols introduces pronounced shifts in pathway dominance and reversibility. Ethanol-blended systems (C3, C5, C4, C6) continue to feature $NO_2$–$HNO_3$ cycling, particularly at 2000 K, where $NO_2 \rightleftharpoons HNO_3$ remains among the top reversible pathways. This suggests ethanol's role in temporarily stabilizing $NO_2$ into higher-order oxides like $HNO_3$, a phenomenon that contributes to reduced free NO concentrations. At 3000 K, ethanol-enhanced systems (notably C6) show elevated activity in $NO \rightleftharpoons HNO_2$ and $HNO_2 \rightleftharpoons NO+NO$ reactions, reinforcing the interpretation that ethanol increases cycling efficiency within intermediate species, thus delaying direct NO buildup.

In contrast, methanol-enriched systems (C7, C9 at 2000 K and C8, C10 at 3000 K) reveal a marked shift in reaction priorities. The most frequent reactions are centred on $NO \rightleftharpoons HNO$ and $NO \rightleftharpoons HNO_2$, with strong reverse reaction rates that exceed or match forward activity. This indicates methanol's robust effect in diverting reactive nitrogen intermediates away from $NO_2$ production. Notably, in C10, the $NO \rightleftharpoons HNO_2$ reverse flux reaches 46, among the highest of all cases, highlighting methanol's superior capacity to suppress $NO_2$ accumulation via intermediate stabilization. These trends are consistent with methanol's high oxygen content and rapid radical propagation, which facilitate hydrogen transfer reactions that reduce NO directly to inert or less reactive species.

The results presented in Table 2 demonstrate that both ethanol and methanol additives effectively reconfigure the $NO_x$ reaction pathways, albeit via distinct chemical routes. Ethanol fosters temporary sequestration of $NO_2$ into $HNO_3$ under mild conditions and supports rapid $NO$–$NO_2$ cycling under high temperatures. Methanol, by contrast, emphasizes intermediate suppression and pathway deflection, preventing NO from converting into $NO_2$ or higher-order $NO_x$ species.

Table 2. Dominant $NO_x$ reaction pathways and their forward and reverse frequencies across C1–C10 fuel systems.

| Reaction \ Systems | Forward Reaction Frequency | | | | | | | | | |
|---|---|---|---|---|---|---|---|---|---|---|
| | C1 | C2 | C3 | C4 | C5 | C6 | C7 | C8 | C9 | C10 |
| NO2 ⇌ HNO3 | 64 | 0 | 50 | 0 | 52 | 0 | 52 | 0 | 58 | 0 |
| NO ⇌ HNO | 0 | 0 | 0 | 36 | 0 | 42 | 0 | 42 | 0 | 28 |
| NO ⇌ HNO2 | 0 | 27 | 0 | 9 | 8 | 42 | 0 | 36 | 0 | 48 |
| NO2 ⇌ NO | 14 | 16 | 26 | 22 | 18 | 20 | 36 | 24 | 20 | 22 |
| NO + NO ⇌ HNO2 | 0 | 20 | 0 | 16 | 0 | 0 | 0 | 39 | 0 | 24 |
| HNO2 ⇌ NO | 0 | 21 | 6 | 0 | 0 | 0 | 14 | 0 | 12 | 0 |
| HNO3 ⇌ NO2 | 0 | 0 | 0 | 0 | 18 | 0 | 0 | 0 | 0 | 0 |
| NO ⇌ N#N | 0 | 6 | 0 | 8 | 0 | 10 | 0 | 9 | 0 | 4 |
| HNO2 + HNO ⇌ NO + NO | 0 | 0 | 0 | 0 | 0 | 12 | 0 | 16 | 0 | 20 |
| | Reverse Reaction Frequency | | | | | | | | | |
| NO2 ⇌ HNO3 | 46 | 0 | 48 | 0 | 48 | 0 | 50 | 0 | 52 | 46 |
| NO ⇌ HNO2 | 0 | 26 | 0 | 22 | 6 | 30 | 0 | 32 | 0 | 46 |
| NO ⇌ HNO | 0 | 12 | 0 | 28 | 0 | 44 | 0 | 30 | 0 | 20 |
| NO ⇌ NO2 | 12 | 14 | 10 | 16 | 14 | 18 | 16 | 14 | 6 | 12 |
| HNO3 ⇌ NO2 | 0 | 0 | 0 | 0 | 9 | 0 | 20 | 0 | 0 | 0 |
| HNO2 ⇌ NO | 0 | 15 | 4 | 0 | 0 | 36 | 4 | 0 | 10 | 0 |
| NO2+NO ⇌ HNO2 | 3 | 15 | 0 | 0 | 0 | 0 | 0 | 18 | 0 | 3 |
| NO ⇌ N#N | 0 | 2 | 0 | 6 | 0 | 6 | 0 | 0 | 0 | 4 |
| NO ⇌ HNO2+HNO | 0 | 9 | 3 | 9 | 0 | 0 | 0 | 15 | 0 | 0 |
| HNO2+HNO ⇌ NO+NO | 3 | 12 | 0 | 4 | 9 | 8 | 4 | 12 | 4 | 16 |

3.2. ML predictions

This section explores how ML models are applied to predict NOx emissions from ammonia–methane–alcohol combustion based on previous MD simulation data. The goal is twofold: we first conduct model performance assessments with different architectures and then demonstrate the ability to make extrapolative predictions for off-design alcohol blends at 2%, 7%, and 12%. By training on simulation-derived features that characterise the charge distribution, energy dynamics, and thermophysical properties, the ML models were hoped to identify patterns in NOx behaviour without running additional computationally expensive MD simulations. Four algorithms were considered in the predictions, including RFR, GBR, SVR, and FCNN, along with four distinct architectural configurations in each model. The comparison of predictions between these models allows for a deeper understanding of how algorithm design affects the predictive accuracy, stability, and feature sensitivity.

3.2.1. Model performance evaluation

The training process utilized 80% of the data while the test set contained the remaining 20%. The evaluation process for all the models utilized five-fold cross-validation to maintain consistency across algorithms while utilizing the root mean square error (RMSE) as the primary measure of scoring performance. For each ML model, four configurations were tested, varying in depth, complexity, and key hyperparameters (e.g., number of estimators for RFR and GBR, regularisation constants for SVR, and neuron layout and dropout for FCNN). Although each algorithm followed the same cross-validation-based selection strategy, Fig. 7 presents the procedure for RFR as a representative example of how model architecture was tuned and evaluated to identify the most reliable configuration. The selection process for Random Forest models is illustrated in Fig. 7, where the performance of four architectures is benchmarked using RMSE across validation folds. The RFR2 configuration emerged as the best performer, offering the lowest error and most stable training-to-test generalisation. Similar procedures were applied to GBR, SVR, and FCNN, with their optimal architectures, e.g., GBR4, SVR1, and FCNN2, which were selected on the same basis. The superior performance of RFR2 can be attributed to its architectural depth and optimal tuning of hyperparameters such as the number of estimators (600) and controlled tree depth (maximum depth of 20), which provided a robust bias–variance balance. Unlike SVR1 and FCNN2, which struggled with the complex nonlinearities intrinsic to combustion chemistry and MD data, RFR2 benefited from ensemble averaging across numerous trees, effectively capturing the stochastic and thermochemical variability of MD-derived features. In contrast, SVR1, while demonstrating reliable performance under constrained input dimensions, failed to generalise across multi-variable interactions. Particularly when trained with larger batch sizes (128), FCNN2, which employed a deep swish-activated 5-layered architecture, performed well, suggesting its use for highly nonlinear problems with dense feature entanglement. Low learning rates and regularisation strategies helped the tree-based GBR4 to perform remarkably well. RFR2, on the other hand, confirmed its place as the best overall performer in the commotion of NOx prediction by reconciling noise, capturing high-dimensional nonlinearity, and providing interpretable variable importances.

Quantitative comparisons for all models are summarised in Table 3, which reports the MMAE, MSE, PCC, and $R^2$ values for each configuration. Among all models, RFR delivered superior performance, showing MAE and MSE values significantly close to zero and $R^2$ and PCC measurements close to 1. SVR showed higher variance and lower accuracy because it struggled to detect complex non-linear relationships in the feature space despite GBR and FCNN presenting respectable results. The results demonstrate the effectiveness of ensemble models such as RFR when dealing with complex and interrelated features extracted from atomistic simulations. The RFR2 outperformed all the other models because it had the lowest MAE (0.661) and highest $R^2$ (0.993), indicating superior predictive capability. Strong performance from the FCNN models, as their MAE scores ranged from 1.740 to 2.036 and their $R^2$ scores peaked at 0.940, showing consistent and generally correct predictions. The GBR demonstrated moderate performance (e.g., GBR3). The SVR exhibited the lowest performance level as shown by its high error margin (MAE = 2.285) and relatively poor R² value (0.903), indicating constrained generalization capacity. The results demonstrate RFR's strong performance capabilities and establish FCNN as a powerful substitute specifically when dealing with intricate feature interactions. The combination of their robustness with their low sensitivity to noise made these models ideal for the prediction of NOx emissions in chemical reactions. The following discussions examine both the learning process of the four models from simulation data and their ability to predict NOx emissions for unknown alcohol concentrations along with the physical importance of key features.

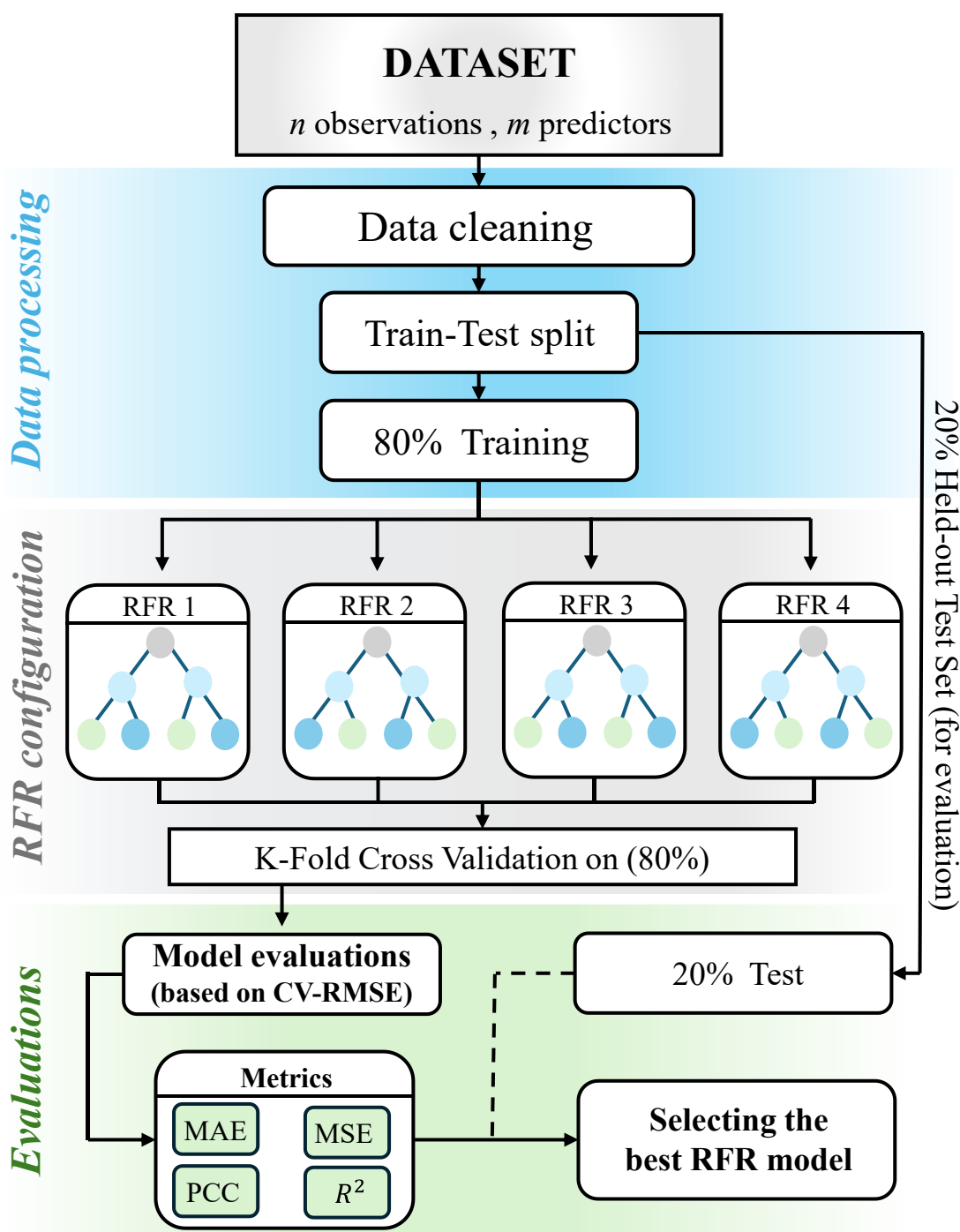

**Fig. 7** Workflow of Random Forest Regression (RFR) model development including data processing, training of four configurations (RFR 1–4), and evaluation using multiple metrics to select the best-performing model.

**Table 3** Model comparison of NOx (ppm) prediction between various ML models (RFR, FCNN, GBR, and SVR) in terms of evaluation metrics (MAE, MSE, PCC, and R²).

| Output | Model | MAE | MSE | PCC | R² |
|---|---|---|---|---|---|
| NOx (ppm) | FCNN 1 | 2.036 | 6.616 | 0.960 | 0.919 |
| | FCNN 2 | 1.740 | 4.913 | 0.970 | 0.940 |
| | FCNN 3 | 1.847 | 5.539 | 0.967 | 0.932 |
| | FCNN 4 | 1.824 | 5.359 | 0.967 | 0.935 |
| | GBR 1 | 2.299 | 8.394 | 0.947 | 0.897 |
| | GBR 2 | 2.100 | 7.641 | 0.952 | 0.912 |
| | GBR 3 | 1.953 | 7.001 | 0.961 | 0.922 |
| | GBR 4 | 1.849 | 6.488 | 0.966 | 0.925 |
| | RFR 1 | 1.012 | 1.715 | 0.990 | 0.983 |
| | RFR 2 | 0.661 | 0.764 | 0.995 | 0.993 |
| | RFR 3 | 0.798 | 1.105 | 0.993 | 0.991 |
| | RFR 4 | 0.684 | 0.833 | 0.996 | 0.979 |
| | SVR 1 | 2.285 | 7.941 | 0.951 | 0.903 |
| | SVR 2 | 2.295 | 7.932 | 0.951 | 0.903 |
| | SVR 3 | 2.285 | 7.942 | 0.951 | 0.903 |
| | SVR 4 | 2.308 | 7.842 | 0.951 | 0.904 |

As shown in Fig. 8, a comparison of predicted NOx values by the best models from each ML group against the actual values has been made. The scatterplots demonstrate the correspondence between model outputs and real data (MD data) values while the dashed diagonal line indicates perfect prediction accuracy. Among all models, RFR demonstrates minimal divergence from the 1:1 line, indicating its exceptional capability to model nonlinear relationships within the combustion data. The SVR1 (Fig. 8(a)) and FCNN2 (Fig. 8(b)) show a wider dispersion that demonstrates significant prediction errors mainly at both ends of the NOx range. This pattern reinforces earlier numerical results: The tree-based RFR2 model demonstrates superior robustness in capturing feature interactions and achieves better generalization across different alcohol compositions. FCNN2 demonstrates adequate predictive capability but shows signs of overfitting or marginal case learning deficiencies due to increased scatter at elevated NOx levels. SVR1 demonstrates effective trend approximation capabilities but suffers from slight bias and reduced performance at extreme data points. These trends are consistent with the values of MAE, MSE, and R² reported in Table 3, further confirming the statistical superiority of RFR2 model.

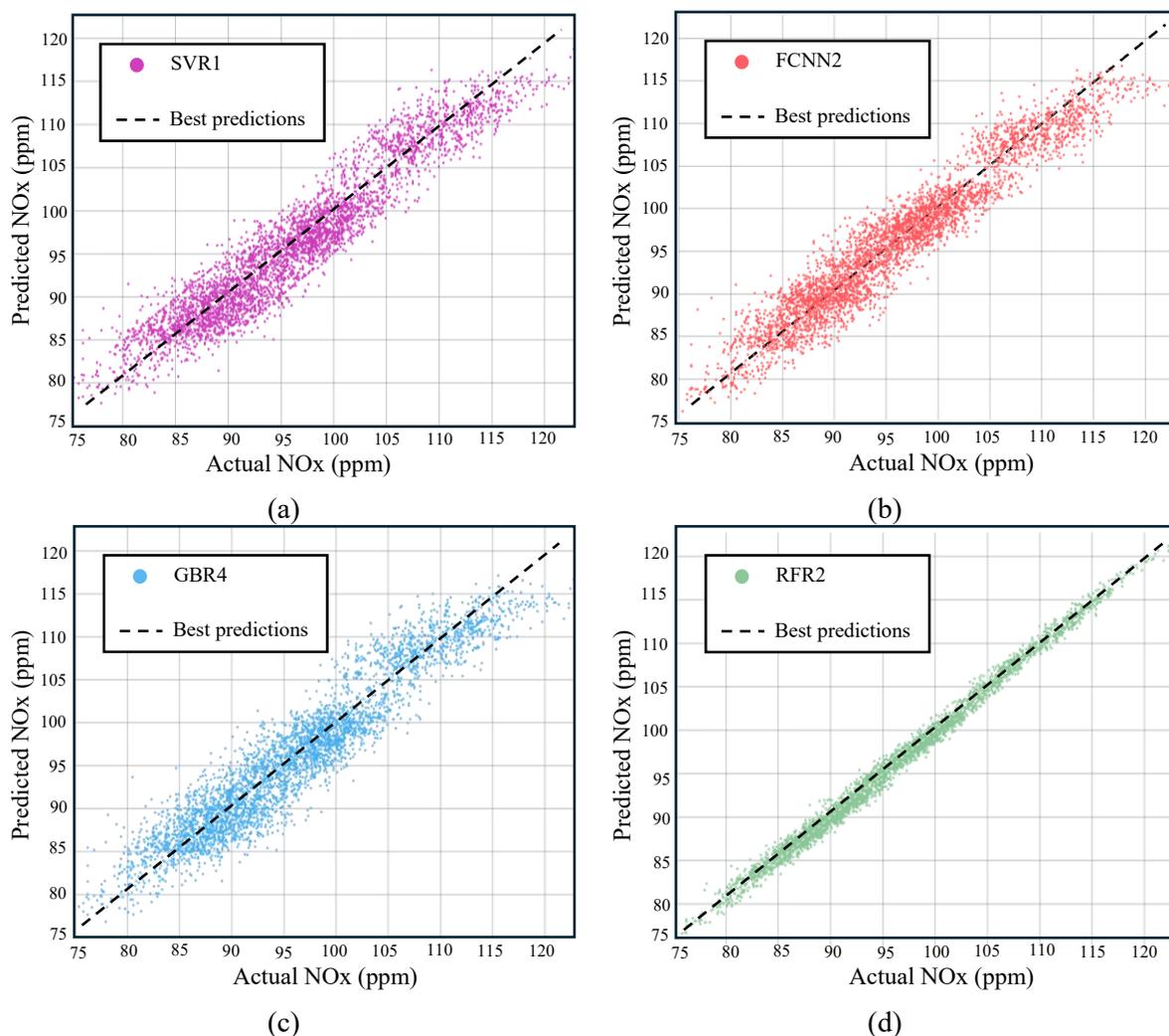

**Fig. 8** Actual vs. predicted NOx (ppm) values for the best configuration of each ML model (RFR2, GBR4, SVR1, and FCNN2): (a) SVR1; (b) FCNN2; (c) GBR4; and (d) RFR2. RFR2 shows the closest match to the ideal prediction ($y = x$), indicating the best performance. SVR exhibits the highest deviation.

Turning to learning curves in Fig. 9, additional insight is gained into the stability and training behaviour of each model. RFR2 again leads with the lowest test set error and smooth convergence across iterations, indicating both reliability and minimal overfitting. Its error bars shrink progressively, showing increased confidence with additional training data. In comparison, FCNN2's learning pattern is erratic along with training error fluctuating between epochs, suggesting sensitivity to architecture or potential instability during optimisation. This behaviour is common in deep learning models when hyperparameters are not fully tuned or the dataset lacks sufficient diversity to support deeper generalisation. GBR4 and SVR1 models follow expected paths: GBR4 progresses through gradual convergence whereas SVR1 shows a significant difference between its training and test performance, hinting potential underfitting or improper regularisation. SVR1's performance gap between training and test outcomes underscores its inability to manage intricate, high-dimensional data. These plots serve not only to validate the ranking of model performance but also to visualise the chemical features behind those outcomes. Although the ensemble learning structure of RFR2 shows optimal performance with the combustion dataset, FCNN2 needs additional adjustments to achieve comparable results along with its nonlinear learning capabilities.

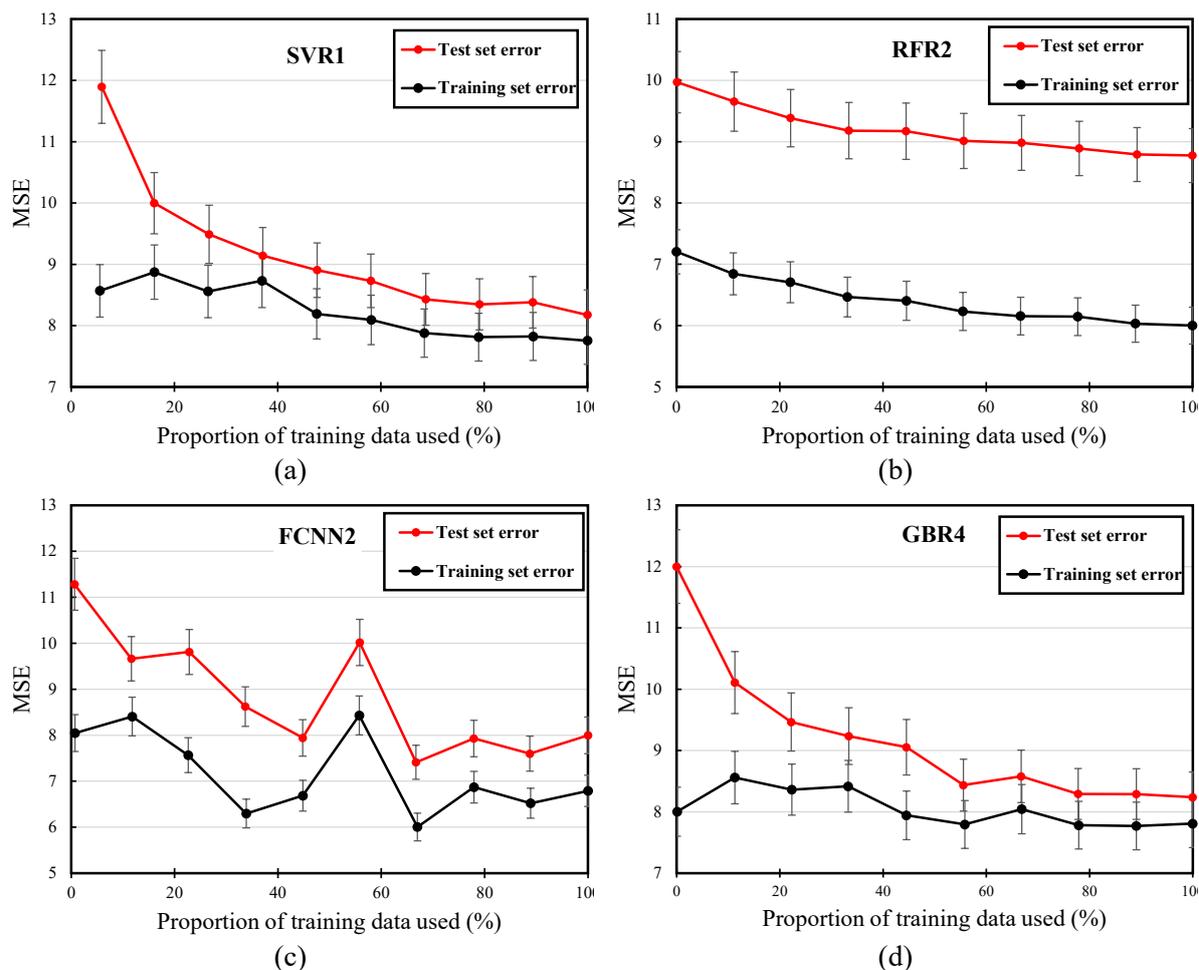

**Fig. 9** Learning curves for the best-performing ML models. Plots show training and test RMSE across training epochs (FCNN) or estimators (RFR/GBR), highlighting convergence behaviour and model stability.

3.2.2. Validation of ML predictions

The predictive robustness of trained ML models was assessed by comparing their outputs against MD-derived NOx values over time across ten combustion cases (C1–C10) that examined different alcohol concentrations (0%, 5%, 10%) and both ethanol and methanol blends at temperatures of 2,000 K and 3,000 K. Figs. 10 (a–j) presents the time-resolved predictions for NOx (ppm) from four ML models, e.g., RFR (yellow), FCNN (red), GBR (sky blue), and SVR (dark blue), alongside the MD reference results (black lines).

Across all systems, RFR consistently yielded the closest agreement with MD simulations, closely tracking both the peak intensities and the short-term fluctuations. This fidelity was particularly evident in ethanol cases such as C3 and C5 (5% and 10% ethanol shown in Figs. 10(c) and (e), where RFR closely captured the onset of NO formation and subsequent oscillatory behaviour. FCNN, benefiting from deep learning architecture and multi-layer abstraction, also demonstrated high predictive accuracy at both low and high temperature scenarios, with performance standing as the second place only after RFR. For example, in C2 (0% Alc. at 3,000 K in Fig. 10(b)), FCNN closely followed MD dynamics during both the peak and relaxation phases. SVR and GBR, by contrast, showed significant deviation in most cases. SVR often underpredicted rapid transitions (e.g., C7, 5% methanol in Fig. 10(f)), while GBR exhibited over-smoothing and lagged behind MD trends, failing to handle the occasional NOx shifts in methanol-rich systems such as C10 (see Fig. 10(j)). The ensemble method used in RFR enables real-time, adaptable responses to sudden changes, showcasing vital capacity of monitoring transient combustion chemistry. FCNN is ideal for systems with complex alcohol–radical interactions because it can model highly intricate non-linear relationships. In contrast, SVR's fixed kernel design and GBR's reliance on a sequence of shallow learners limit their adaptability in fast-changing chemical environments such as the rapid NOx fluctuations occurring during combustion.

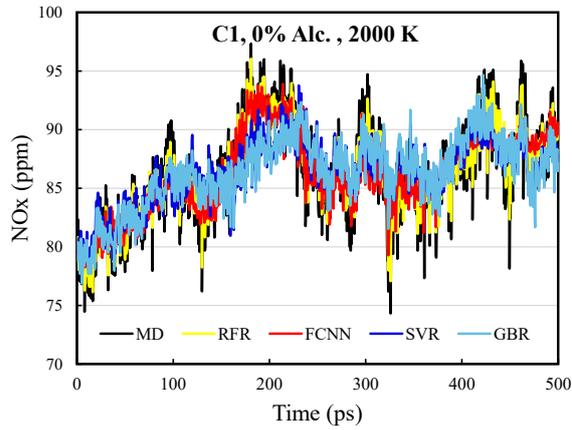

(a)

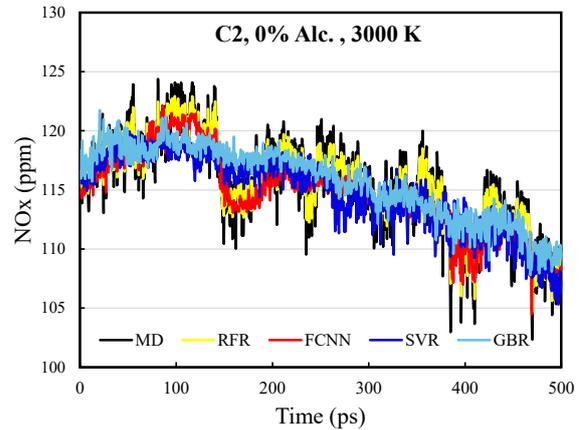

(b)

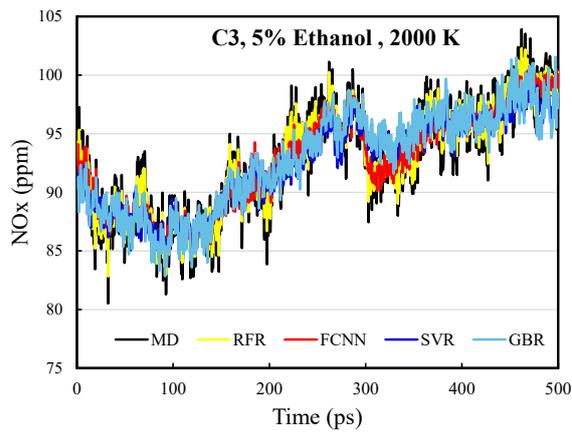

(c)

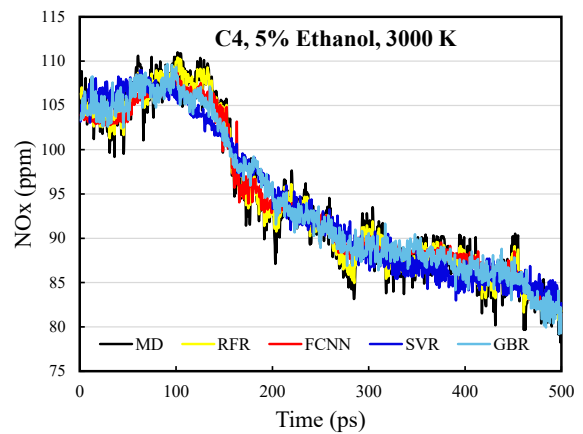

(d)

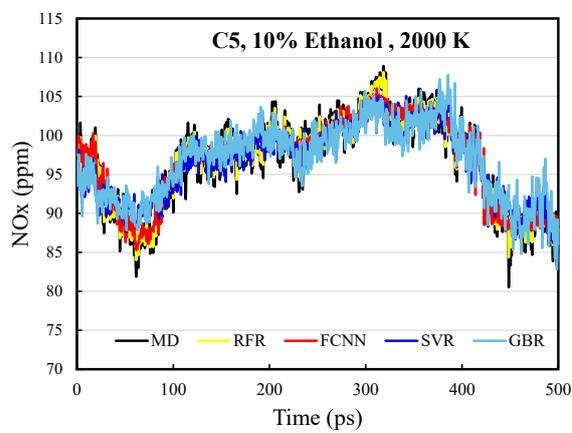

(e)

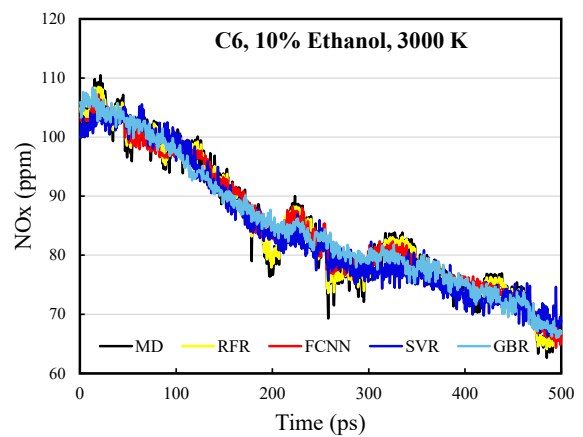

(f)

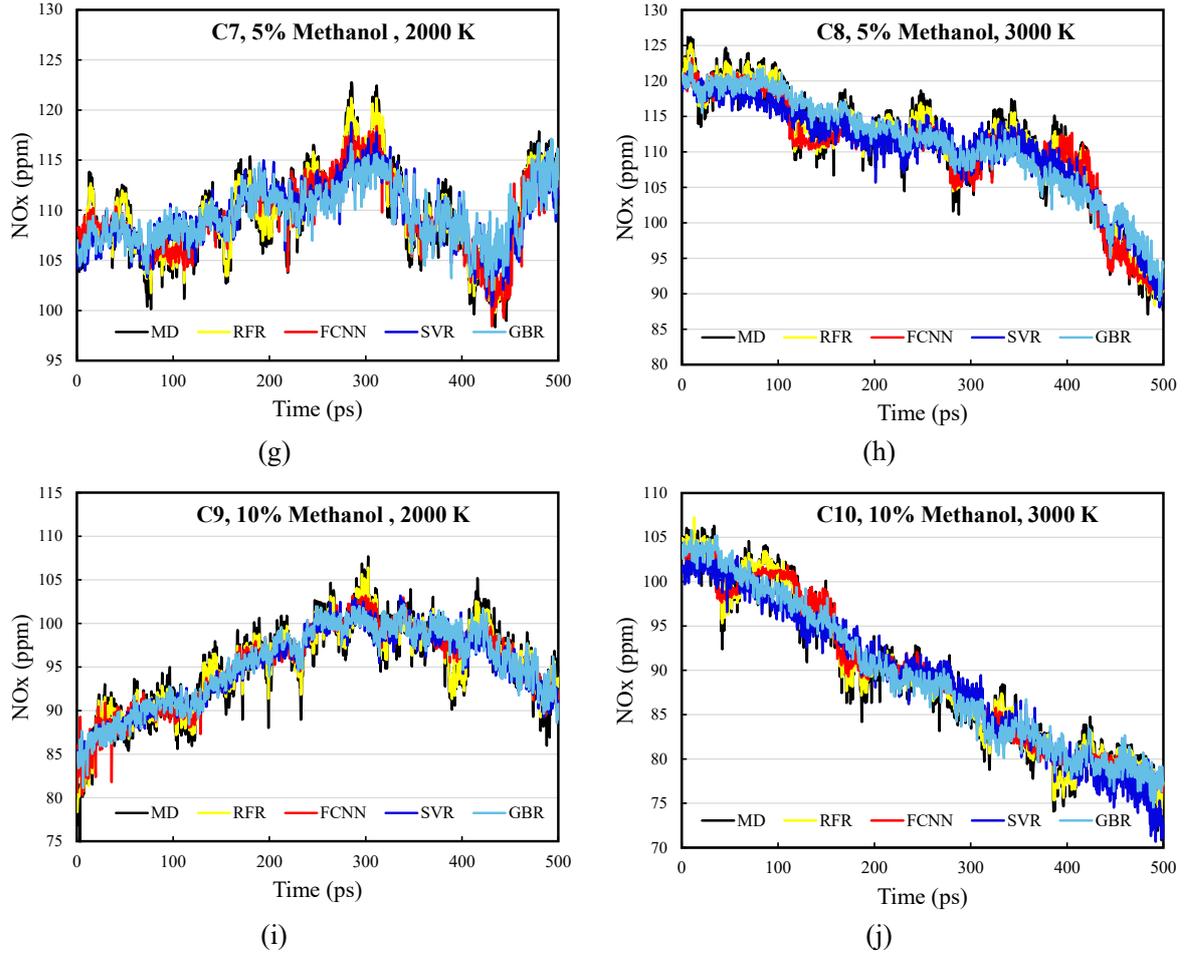

**Fig. 10** Comparison of NOx evolution predicted by four ML models (e.g., RFR, FCNN, SVR, and GBR) against ReaxFF MD data for 10 combustion cases (C1–C10) at 2,000 K and 3,000 K: (a) C1 at 2,000 K; (b) C2 at 3,000 K; (c) C3 at 2,000 K; (d) C4 at 3,000 K; (e) C5 at 2,000 K; (f) C6 at 3,000 K; (g) C7 at 2,000 K; (h) C8 at 3,000 K; (i) C9 at 2,000 K; and (j) C10 at 3,000 K.

Tables 4 and 5 display a direct comparison of NOx emissions between ML model predictions and MD simulation results in the base fuel systems, C1 at 2,000 K and C2 at 3,000 K, which complement the previously discussed performance metrics and learning curves. The analysis reports ten separate time steps ranging from 50 to 500ps to consider the complete trajectory span. The RFR model demonstrates superior accuracy through its lowest recorded average absolute and relative errors at every time step. RFR exhibited the lowest error margins, with percentage errors consistently below 1.6% at 2,000 K (see Table 4) and under 1% for most time steps at 3,000 K (see Table 5). For instance, RFR predicted 86.896ppm vs. 87.883ppm in MD at 250ps, reporting an absolute error of only 0.987ppm (1.12%). FCNN also maintained strong accuracy with errors generally under 2.5%, occasionally outperforming RFR at later time steps, such as at 500ps in C2 in Table 5, where it only had an error of 0.14ppm. SVR and GBR showed good performance but displayed minor over- or under-predictions of NOx during longer reaction periods (e.g., 300–500ps) which aligns with findings from predicted–actual plots (Figs. 8) and sensitivity analysis (Fig. 11). For example, at 300ps in C1 in Table 4, the reported error by SVR reached 5.707ppm (6.15%), which was closely followed by GBR. This quantitative evaluation supports the observed trends from the performance plots specifically the actual vs. predicted correlations in Fig. 8, the learning curves in Fig. 9, and the NO$_x$ evolution patterns predicted by the four ML models, as shown in Fig. 10, reinforce the established performance hierarchy: RFR2 > FCNN2 > GBR4 ≈ SVR1., with RFR2 and FCNN2 emerging as the most effective configurations (see Table 5). While the results align well with the behaviour observed in physics-based simulations, this consistency suggests that machine learning models—particularly ensemble approaches—can capture key patterns in NOx evolution within ammonia–methane combustion systems when trained appropriately

3.2.3. Sensitivity analysis of key predictors

To deepen understanding of the ML models' decision-making processes, a sensitivity analysis was performed on the most influential predictors contributing to NOx emission predictions. These predictors were selected based on their normalised importance scores within the best-performing architecture of each model: RFR2, SVR1, GBR4, and FCNN2 (see Fig. 11). It should be noted that only the most impactful variables are shown, not the full feature set of 26 inputs. The RFR2 model identifies Timestep along with PotEng (potential energy) and v_ecoa (valence angle conjugation energy) as the leading three contributors at 2,000 K, followed by v_qC (partial charge on C atoms). Timestep and PotEng are physically intuitive parameters because they represent the reactive system's dynamic and thermodynamic evolution. The v_qC measurement shows how C atoms participate in oxidation reactions with OH and O radicals when alcohol additives are present. The RFR model's internal weighting aligns with known chemical kinetics, showing its high interpretability and reliability. At 3,000 K, RFR2 places even greater weight on TotEng (total system energy), v_qO (partial charge on O atoms), and v_eb (bond energy), all of which are strongly correlated with high-temperature reactivity. The appearance of v_ep (Coulombic interaction energy) and v_eqeq (charge equilibration energy) further confirms that charge redistribution becomes a dominant mechanism in NOx formation under elevated temperatures, which is consistent with conclusions supported by MD results in Sec. 3.1. For GBR4 model, the pattern is similar but places more emphasis on electron-specific descriptors. At 2,000 K, v_ecoa, Timestep, and both alcohol types (methanol and ethanol) play leading roles. At 3,000 K, TotEng, v_qC, and mass fraction of alcohol are primary, followed by v_ecoa, v_eco (conjugation energy), and v_ca (atom-centred charge energy). These findings suggest that GBR captures both global thermodynamic trends and local charge effects, despite corresponding less direct relation to combustion chemistry compared with RFR2. The SVR1 model, while achieving decent performance, relies more heavily on high-level variables such as methanol, mass fraction of alcohol, and Timestep, which are the most influentialat 2,000 K. At 3,000 K, the shift toward Timestep, Density, and TotEng shows that SVR1 responds more to global system states than fine-grained chemical details. Notably, features such as v_eqeq, v_qC, and v_ew (van der Waals energy) also appear, indicating that SVR1 can detect some electronic effects, though less consistently. The FCNN2 model, representing a deep learning approach, reveals a distinct pattern of feature prioritization. At 2,000 K, the most influential features include Methanol and Alcohol_% dominate, followed by v_ecoa, Timestep, and v_qC. At 3,000 K, the model assigns highest importance to Timestep, Alcohol_%, v_qC, PotEng, and lower-level features such as v_ep and v_elp (lone-pair energy). FCNN2 appears to focus more on general system reactivity and alcohol input terms, potentially due to its capacity to encode nonlinear relationships across broad feature spaces. This aligns with typical deep learning behaviour, where low-level patterns are distributed across multiple internal layers rather than interpreted individually. Among all models, RFR demonstrates the strongest alignment between its predictive logic and the underlying combustion mechanisms. Its preference for chemically meaningful variables such as v_qO, v_qC, PotEng, and v_ecoa reinforces its status as the most chemically interpretable model in this study. Its ability to shift emphasis from carbon-related dynamics at 2,000 K to oxygen and total energy at 3,000 K reflects the actual behaviour of NOx precursors and formation pathways observed in MD simulations.

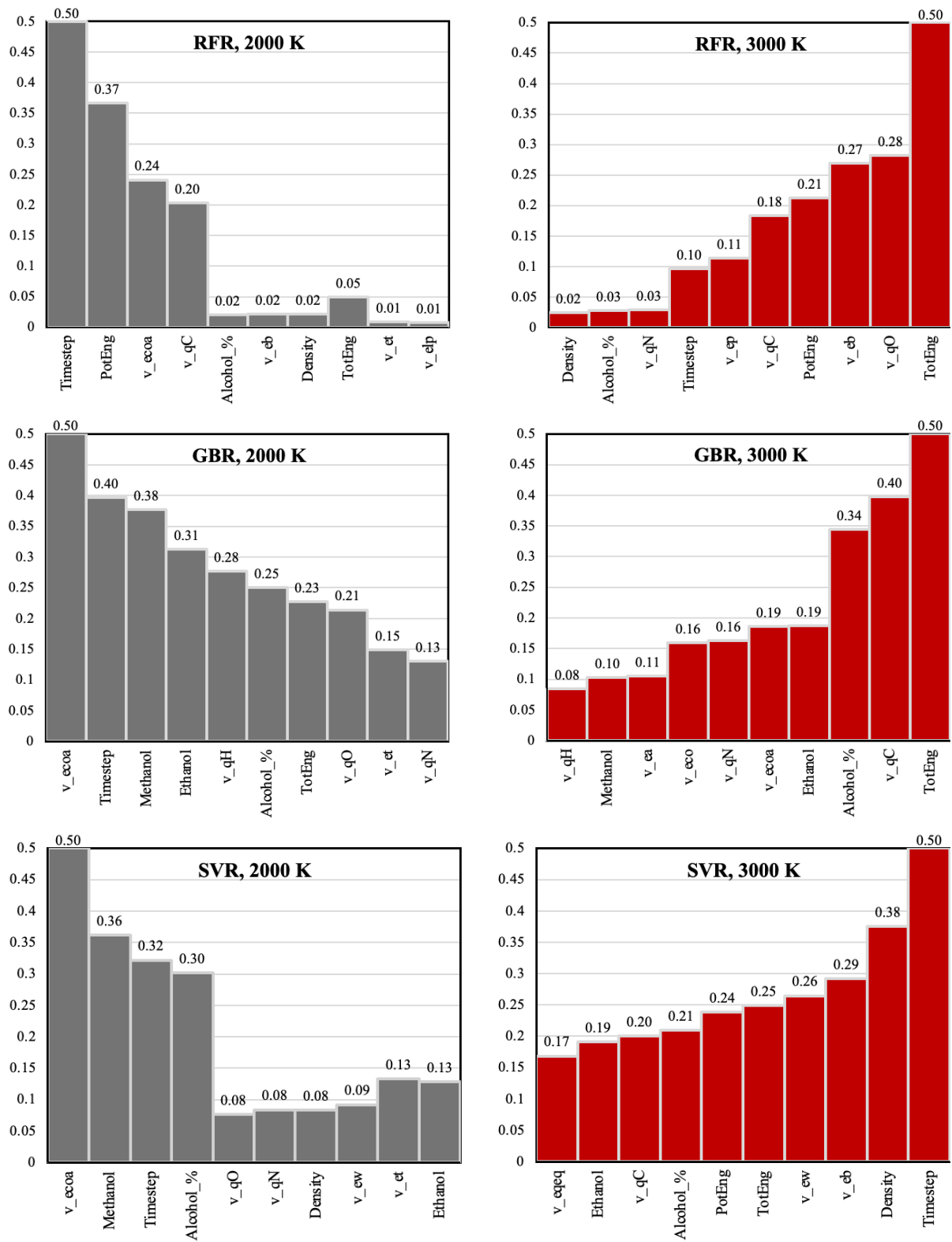

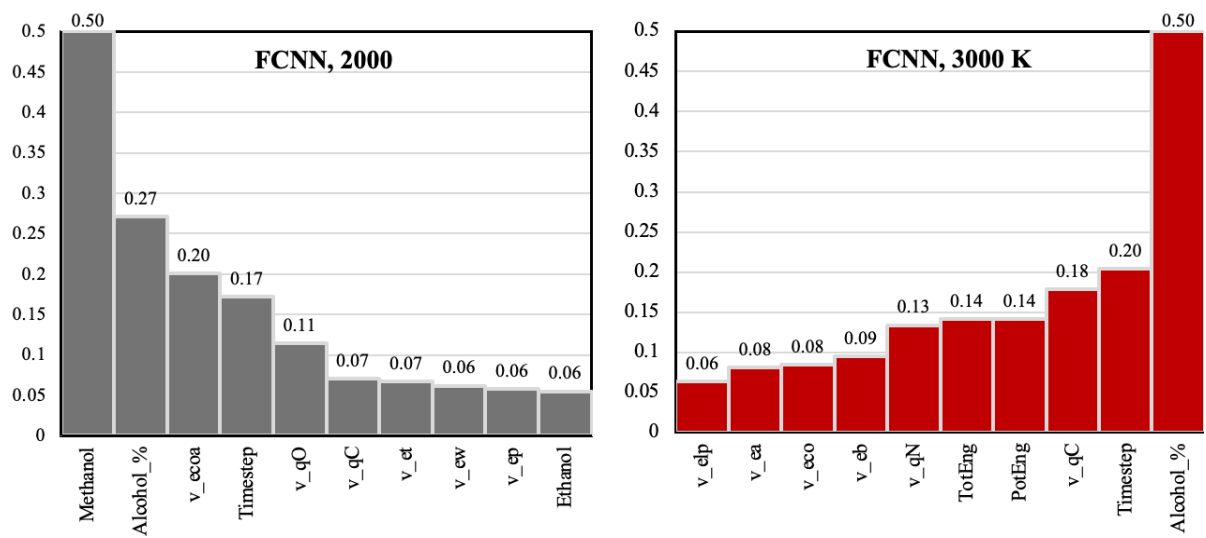

**Fig. 11** Sensitivity analysis of the most influential input features for NOx prediction among the best-performing models (RFR2, GBR4, SVR1, and FCNN2) at 2,000 K and 3,000 K. Variable importance values are normalised to highlight the key contributions from physical, chemical and electronic descriptors.

**Table 4** Comparison of NOx emissions between ML model predictions and MD simulation results at 2,000 K (System C1, 0% Alcohol).

| Output | System | Temp. | Time (ps) | MD | RFR | RFR Abs. error | RFR percentage error | FCNN | FCNN Abs. error | FCNN percentage error | SVR | SVR Abs. error | SVR percentage error | GBR | GBR Abs. error | GBR percentage error |
|---|---|---|---|---|---|---|---|---|---|---|---|---|---|---|---|---|
| NOx (ppm) | C1 | 2000 K | 50 | 77.852 | 78.892 | 1.040 | 1.336 | 82.643 | 4.791 | 6.154 | 82.103 | 4.251 | 5.460 | 81.592 | 3.739 | 4.803 |
| | | | 100 | 89.812 | 85.155 | 4.658 | 5.186 | 85.066 | 4.746 | 5.284 | 85.003 | 4.809 | 5.355 | 85.260 | 4.552 | 5.068 |
| | | | 150 | 86.782 | 86.584 | 0.198 | 0.228 | 85.227 | 1.555 | 1.792 | 84.225 | 2.558 | 2.947 | 84.381 | 2.401 | 2.767 |
| | | | 200 | 91.031 | 91.582 | 0.552 | 0.606 | 92.155 | 1.124 | 1.235 | 90.330 | 0.700 | 0.769 | 90.757 | 0.274 | 0.300 |
| | | | 250 | 87.883 | 86.896 | 0.987 | 1.123 | 85.058 | 2.824 | 3.214 | 86.061 | 1.822 | 2.073 | 86.515 | 1.367 | 1.556 |
| | | | 300 | 92.867 | 87.341 | 5.526 | 5.951 | 86.412 | 6.455 | 6.951 | 87.160 | 5.707 | 6.145 | 88.286 | 4.581 | 4.933 |
| | | | 350 | 87.719 | 86.873 | 0.846 | 0.965 | 84.628 | 3.091 | 3.524 | 86.669 | 1.050 | 1.197 | 86.440 | 1.279 | 1.458 |
| | | | 400 | 89.310 | 88.941 | 0.369 | 0.413 | 87.765 | 1.545 | 1.730 | 87.570 | 1.740 | 1.948 | 88.226 | 1.084 | 1.213 |
| | | | 450 | 83.888 | 85.239 | 1.350 | 1.610 | 88.963 | 5.074 | 6.049 | 87.737 | 3.849 | 4.588 | 87.896 | 4.008 | 4.778 |
| | | | 500 | 90.788 | 90.189 | 0.599 | 0.660 | 90.576 | 0.212 | 0.233 | 87.893 | 2.895 | 3.189 | 87.107 | 3.680 | 4.054 |

**Table 5** Comparison of NOx emissions between ML model predictions and MD simulation results at 3,000 K (System C2, 0% Alcohol).

| Output | System | Temp. | Time (ps) | MD | RFR | RFR Abs. error | RFR Percentage error | FCNN | FCNN Abs. error | FCNN Percentage error | SVR | SVR Abs. error | SVR Percentage error | GBR | GBR Abs. error | GBR Percentage error |
|---|---|---|---|---|---|---|---|---|---|---|---|---|---|---|---|---|
| NOx (ppm) | C2 | 3000 K | 50 | 99.89 | 99.84 | 0.04 | 0.04 | 99.66 | 0.23 | 0.23 | 99.22 | 0.66 | 0.66 | 102.68 | 2.80 | 2.80 |
| | | | 100 | 103.06 | 101.62 | 1.43 | 1.39 | 101.24 | 1.81 | 1.76 | 97.45 | 5.61 | 5.44 | 99.62 | 3.44 | 3.34 |
| | | | 150 | 97.75 | 97.00 | 0.76 | 0.78 | 96.53 | 1.22 | 1.25 | 94.05 | 3.70 | 3.78 | 95.60 | 2.15 | 2.20 |
| | | | 200 | 92.03 | 91.48 | 0.55 | 0.60 | 90.54 | 1.49 | 1.62 | 91.18 | 0.85 | 0.93 | 91.50 | 0.54 | 0.58 |
| | | | 250 | 89.40 | 89.41 | 0.01 | 0.01 | 90.48 | 1.08 | 1.21 | 89.92 | 0.51 | 0.57 | 87.61 | 1.80 | 2.01 |
| | | | 300 | 83.70 | 85.84 | 2.14 | 2.56 | 85.90 | 2.19 | 2.62 | 87.64 | 3.94 | 4.71 | 84.57 | 0.87 | 1.04 |
| | | | 350 | 83.06 | 82.84 | 0.22 | 0.26 | 81.85 | 1.21 | 1.45 | 82.53 | 0.53 | 0.64 | 83.96 | 0.90 | 1.08 |
| | | | 400 | 78.75 | 78.17 | 0.58 | 0.74 | 79.15 | 0.41 | 0.52 | 79.28 | 0.53 | 0.67 | 78.51 | 0.24 | 0.31 |
| | | | 450 | 80.39 | 79.96 | 0.43 | 0.54 | 79.42 | 0.97 | 1.20 | 75.53 | 4.85 | 6.03 | 77.99 | 2.39 | 2.98 |
| | | | 500 | 76.84 | 77.37 | 0.53 | 0.69 | 76.70 | 0.14 | 0.18 | 72.54 | 4.31 | 5.60 | 77.20 | 0.36 | 0.46 |

## 3.3. Extrapolative predictions for untrained alcohol ratios

Extrapolation, in this context, refers to the model's ability to make reliable predictions for input conditions, here, alcohol concentrations that lie outside the numerical range seen during training. Unlike interpolation, which estimates values between known data points, extrapolation tests whether the model has truly learned the underlying chemical relationships that generalize to unobserved systems. The ability to perform such extrapolative prediction is crucial for real-world combustion research, as it enables low-cost virtual screening of fuel blends without requiring additional expensive and time-consuming simulations. It reflects the model's capacity to generalize beyond known data while still remaining chemically consistent. Fig. 12 shows the ML-predicted NOx profiles for untrained alcohol concentrations, e.g., 2%, 7%, and 12% of both ethanol and methanol at 2,000 K (see Fig. 12(a)) and 3,000 K (see Fig. 12(b)). These systems (C11–C22) were not included in the original training data, meaning they represent a direct test of the model's extrapolation capability. At 2,000 K in Fig. 12(a), the 2% methanol case in C17 yields the highest NOx production, surpassing 110ppm at 500ps. In contrast, 2% ethanol in C15 closely tracks the baseline behaviour of C1, stabilising around 90–95ppm. As concentration increases, this disparity narrows: 7% methanol in C21 maintains higher NOx than 7% ethanol in C19, while at 12% concentration, methanol in C13, grey and ethanol in C11 converge around 95ppm. These results align with earlier MD findings, reaffirming that methanol's aggressive early oxidation at lower temperatures fosters NOx, while ethanol moderates the release of N-containing radicals. The trend shifts greatly at 3,000 K in Fig. 12(b). All systems exhibit reduced NOx formation, but now ethanol-rich blends dominate the suppression. The 12% ethanol case in C12 falls below 70ppm in the end, marking the lowest NOx level among all tested extrapolated cases. Meanwhile, 12% methanol in C14 levels off just under 80ppm. Across mid-level concentrations, 7% ethanol in C20 also outperforms 7% methanol in C22. The low-concentration region again favours ethanol: C16 (2% ethanol) outperforms C18 (2% methanol) by a visible margin.

These trends reinforce two interpretations of the underlying chemical reaction mechanisms. First, methanol is more reactive at low temperatures, likely accelerating $NH_3$ oxidation through earlier radical generation, which in turn drives NO formation. However, as temperature increases, ethanol's slower and more controlled radical propagation favours routes that stabilise intermediates or divert nitrogen away from $NO/NO_2$ loops. Second, the model's ability to recover these patterns, despite having been trained solely on 0%, 5%, and 10% alcohol levels, demonstrates the capacity of chemical-guided ML. Although no MD simulations were run for these off-design systems, the trends match known combustion behaviours, illustrating that the model is capable of capturing real chemical logic and reliable for low-cost predictive screening.

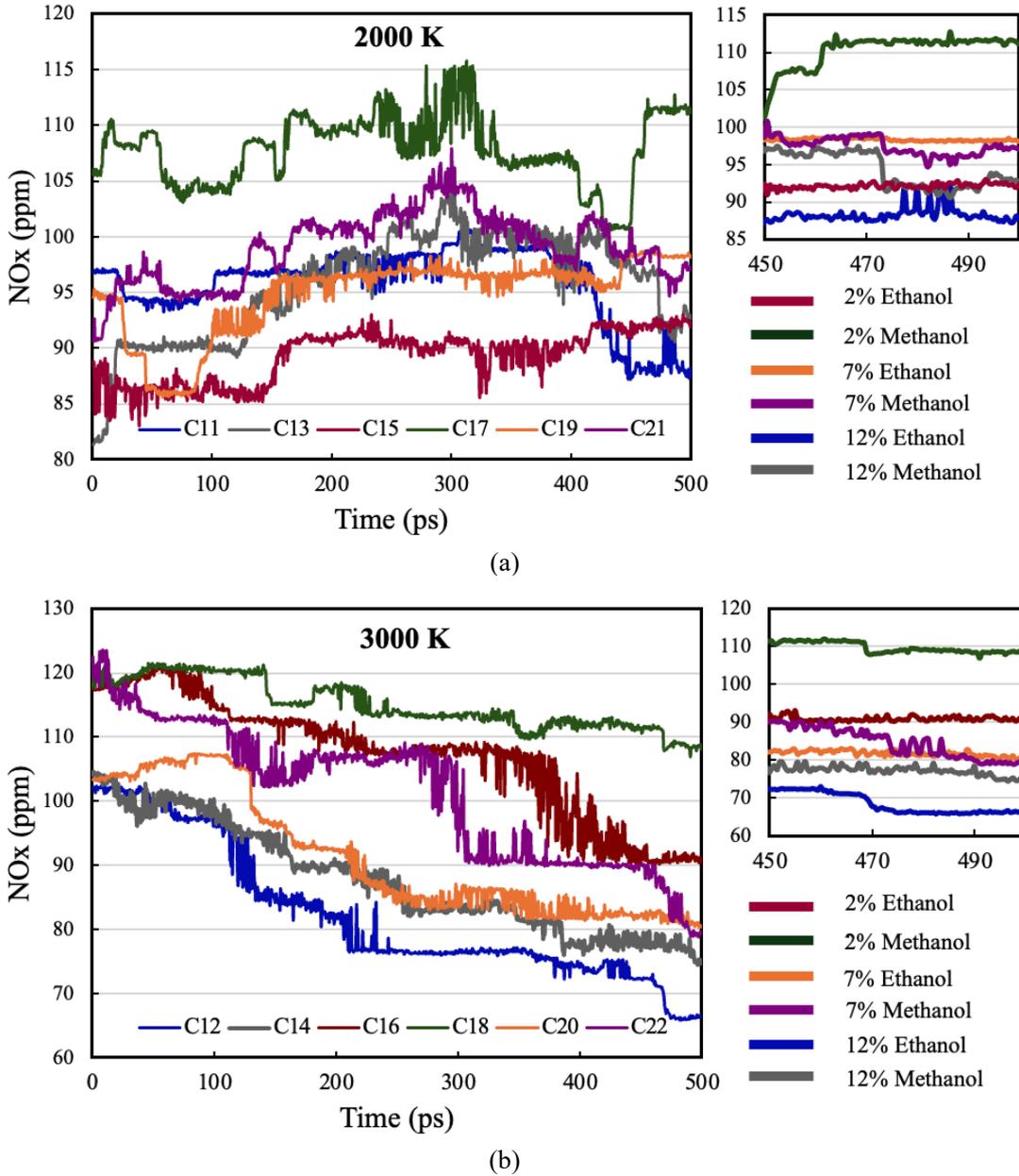

**Fig. 12** ML-predicted NOx emissions for extrapolated alcohol concentrations (2%, 7%, and 12%) of methanol-ethanol in ammonia–methane combustion at: (a) 2,000 K and (b) 3,000 K. The colour codes are: dark red (C15, C16) for 2% ethanol, dark green (C17, C18) for 2% methanol, orange (C19, C20) for 7% ethanol, purple (C21, C22) for 7% methanol, dark blue (C11, C12) for 12% ethanol, and grey (C13, C14) for 12% methanol.

In Tables 6 and 7, predicted values are compared with synthesised ReaxFF MD datasets to validate the extrapolation results for 12% alcohol scenarios. These datasets were predicted for C11, C13 at 2,000 K and C12, C14 at 3,000 K, representing 12% ethanol and 12% methanol, respectively. The results are summarised in terms of absolute and percentage errors for each 100ps period across 500ps. The average percentage error for ethanol is relatively low, increasing from 3.845% at 100ps, to 4.49% at 500ps. The prediction remains stable, but the trend shows an incremental increase in error with time, reflecting the complexity of extrapolation. The percentage error begins at 2.02% when measured at 100ps and increases to 12.78% at 400ps. The rising error trend arises from methanol's increased reactivity, leading to more significant departures from MD results. The percentage error for 3,000 K cases in Table 7 begins at 6.58% at 100ps and decreases to 5.97% at 500ps as the timestep increases. It suggests the model's predictions stabilize as the combustion dynamics approach the steady state at higher temperatures. In addition, a similar error pattern with methanol is observed. The error starts at 3.26% at 100ps but increases significantly to 17.74% at 400ps. The substantial error rise during intermediate timesteps underscores the difficulties in extrapolating results for high methanol concentrations and elevated temperatures. It can be concluded that increased prediction errors in methanol are due to its stronger reactivity that creates complex reaction pathways, making predictive process more challenging for ML models.

**Table 6** Quantitative comparison of RFR model predictions for 12% alcohol concentrations at 2,000 K (cases C11 and C13).

| Output | System | Temp. | Time (ps) | Alc. % | Alc. type | MD | ML | Abs. error | Percentage error |
|---|---|---|---|---|---|---|---|---|---|
| NOx (ppm) | C11 | 2000 K | 100 | 12% | Ethanol | 91.061 | 94.908 | 3.847 | 4.225 |
| | C11 | 2000 K | 200 | 12% | Ethanol | 105.141 | 98.170 | 6.971 | 6.630 |
| | C11 | 2000 K | 300 | 12% | Ethanol | 109.811 | 99.385 | 10.426 | 9.494 |
| | C11 | 2000 K | 400 | 12% | Ethanol | 101.188 | 96.625 | 4.563 | 4.509 |
| | C11 | 2000 K | 500 | 12% | Ethanol | 83.330 | 87.818 | 4.488 | 5.386 |
| | C13 | 2000 K | 100 | 12% | Methanol | 87.706 | 89.726 | 2.020 | 2.303 |
| | C13 | 2000 K | 200 | 12% | Methanol | 88.135 | 95.456 | 7.321 | 8.306 |
| | C13 | 2000 K | 300 | 12% | Methanol | 101.925 | 103.652 | 1.727 | 1.695 |
| | C13 | 2000 K | 400 | 12% | Methanol | 87.059 | 99.838 | 12.779 | 14.679 |
| | C13 | 2000 K | 500 | 12% | Methanol | 88.026 | 92.682 | 4.656 | 5.289 |

**Table 7** Quantitative comparison of RFR model predictions for 12% alcohol concentrations at 3,000 K (cases C12 and C14).

| Output | System | Temp. | Time (ps) | Alc. % | Alc. type | MD | ML | Abs. error | Percentage error |
|---|---|---|---|---|---|---|---|---|---|
| NOx (ppm) | C12 | 3000 K | 100 | 12% | Ethanol | 91.837 | 98.422 | 6.584 | 7.169 |
| | C12 | 3000 K | 200 | 12% | Ethanol | 70.915 | 82.390 | 11.475 | 16.181 |
| | C12 | 3000 K | 300 | 12% | Ethanol | 69.590 | 76.338 | 6.749 | 9.698 |
| | C12 | 3000 K | 400 | 12% | Ethanol | 70.112 | 74.099 | 3.987 | 5.687 |
| | C12 | 3000 K | 500 | 12% | Ethanol | 60.390 | 66.358 | 5.968 | 9.883 |
| | C14 | 3000 K | 100 | 12% | Methanol | 97.138 | 100.400 | 3.262 | 3.359 |
| | C14 | 3000 K | 200 | 12% | Methanol | 83.728 | 90.587 | 6.860 | 8.193 |
| | C14 | 3000 K | 300 | 12% | Methanol | 74.588 | 83.787 | 9.199 | 12.333 |
| | C14 | 3000 K | 400 | 12% | Methanol | 65.553 | 77.181 | 11.628 | 17.738 |
| | C14 | 3000 K | 500 | 12% | Methanol | 71.582 | 74.921 | 3.339 | 4.664 |

Fig. 13 illustrates the mean percent error in ML model predictions for several alcohol concentrations, showing the acceptable margin of error (e.g., ≤ 5% for several cases such as C15, C16, and C18) reflects the robustness of RFR model in predicting NOx emissions for extrapolated cases within reasonable error limits. The ethanol-rich systems exhibit lower average percentage errors, with C11 (12% ethanol at 2,000 K) showing the smallest error (5.33%), while the methanol-rich systems (C13) show the largest error (12.85%) at 2,000 K.

At 3,000 K, the trend persists, with ethanol systems maintaining lower error levels compared with methanol, reflecting ethanol's relatively stable behaviour at higher temperatures. In 2,000 K, the error for methanol tends to increase at higher concentrations, particularly in cases (C17) of 12% methanol. This trend matches what is shown in Tables 6 and 7, where mixtures with methanol have larger errors, both in total amount and percentage, than those with ethanol. The larger errors in methanol-rich systems happen because methanol reacts more in cold conditions, making it harder for models to accurately predict the detailed reaction processes. In contrast, ethanol-rich systems, while still presenting some extrapolative challenges, exhibit more consistent model behaviour, particularly at higher temperatures such as 3000 K.

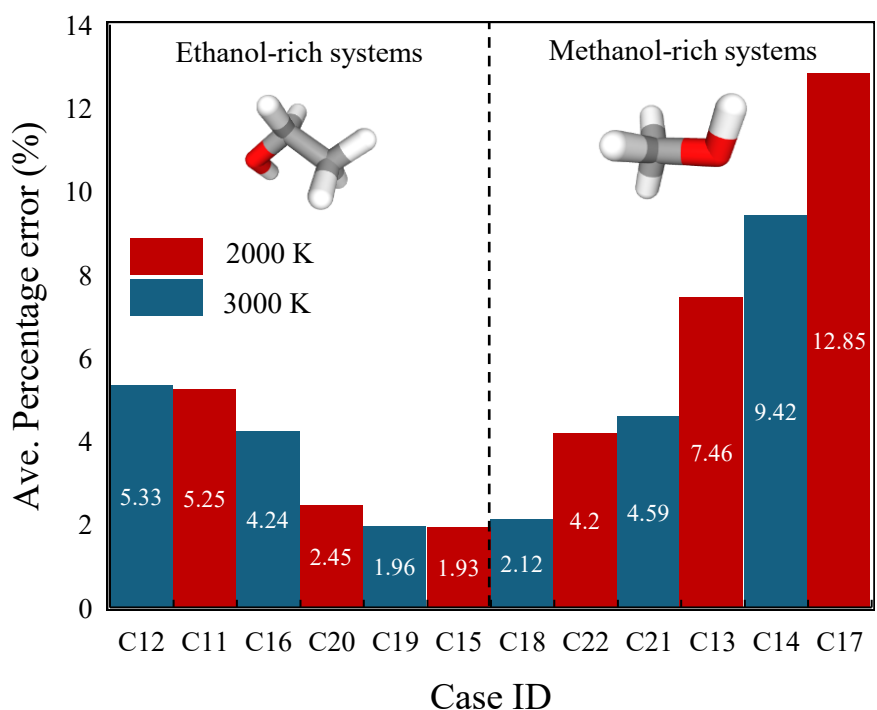

**Fig. 13** Average percentage error of RFR-predicted NOx emissions across extrapolated alcohol concentrations (2%, 7%, and 12%) for different ammonia–methane–alcohol combustion cases (C11-C22) at 2,000 K (red bars) and 3,000 K (blue bars).

While the extrapolation to 12% alcohol is successfully achieved with acceptable accuracy, the extrapolation to higher percentages of alcohol (e.g., 15–20%) is difficult. The increased reactivity and complexity of chemical pathways at higher concentrations make it difficult to maintain high predictive accuracy. When alcohol levels get close to 15-20%, the way the fuel and air mix reacts can change significantly compared with the data used to train the model, leading to bigger differences between what the model predicts and the actual combustion outcomes. This limitation likely stems from a lack of representative training samples capturing the altered charge distributions, energy profiles, and species interaction dynamics unique to these higher alcohol blends. Further work is needed to enhance the extrapolation process, particularly to 15-20% alcohol levels, which may need more training data or better feature chemical reactions to capture non-linear effects at those levels. These non-linearities arise from threshold-driven changes in radical formation, bond dissociation patterns, and secondary reactions that do not scale linearly with alcohol concentration. For example, higher hydroxyl content at ≥15% alcohol may shift dominant NOx formation pathways or trigger new intermediate stabilization loops not seen at lower concentrations.

## 4. Conclusions

This study employed ReaxFF-based molecular dynamics simulations to investigate the impact of ethanol and methanol additives at concentrations of 0%, 5%, and 10% on the combustion behaviour of ammonia–methane mixtures under high-temperature conditions (2,000 K and 3,000 K). The results demonstrated that alcohol enrichment strongly alters charge redistribution, NOx formation, and intermediate stability. At 2,000 K, methanol-rich mixtures (especially C7 and C9) exhibited pronounced electron withdrawal from nitrogen atoms, accelerated $NH_3$ oxidation, and elevated NOx levels, with C7 reaching 110.54 ppm. In contrast, 10% ethanol (C5) suppressed NOx more effectively, producing only 88.31 ppm compared to 90.79 ppm in the base fuel. At 3,000 K, ethanol showed even stronger NOx-reduction performance particularly in C6 (10% ethanol), which achieved the lowest NOx level of 66.46 ppm, reflecting a 39.5% decrease compared to C2 (109.95 ppm). Methanol's performance at high temperature was moderate (C10: 76.84 ppm), due to sustained reactivity and enhanced radical production. Charge equilibration analysis revealed more stable electron distributions in methanol cases at 2,000 K and greater reactivity in ethanol-rich mixtures at 3,000 K. The dominant reaction pathways also shifted: ethanol favoured reversible $NO_2 \rightleftharpoons HNO_3$ loops at low temperatures and nitrate decomposition at high temperatures, while methanol facilitated NO conversion through HNO and $N_2O$ routes. These findings underline that both alcohol type and concentration play pivotal roles in modulating NOx pathways and combustion reactivity. Four regression models were trained with the atomistic descriptors from MD simulations, including Random Forest Regression (RFR), Support Vector Regression (SVR), Gradient Boosting Regression (GBR), and Fully Connected Neural Networks (FCNN). Among these, the RFR2 model (a tree-based ensemble method) achieved the highest accuracy, with a Mean Absolute Error (MAE) of 0.661 and $R^2$ of 0.993, followed by FCNN2 (MAE ≈ 1.74–2.04, $R^2$ = 0.940). GBR4 showed moderate accuracy, while SVR1 had the weakest performance (MAE = 2.285, $R^2$ = 0.903). The RFR model accurately predicted NOx formation within the interpolated range (2%, 5%, 7%, and 10%) as well as for extrapolated alcohol levels (12%), with errors <5% for ethanol-rich mixtures and moderate errors in methanol-rich mixtures. This framework provides a novel predictive route that bypasses the need for direct simulations at every blend ratio, enabling scalable exploration of fuel compositions. Quantitative validation against synthetic MD data showed low prediction errors (≤5%) in most ethanol cases and moderate errors in methanol-rich systems. The results indicate that extrapolation within the 10–12% range is numerically viable, whereas predictions over 15% may necessitate enhanced models or broader datasets to improve the accuracy. It is because the models could not capture the fundamentally altered reaction dynamics and bond dissociation mechanisms observed in that regime. Future work will aim to enhance the extrapolative robustness of the ML models by integrating noise-reduced, chemically valid training data [51] and incorporating reaction-aware physical priors [52] to better capture transition dynamics. In addition, emerging frameworks such as optimal transport for reaction state mapping [53], physics-informed neural networks for reaction–diffusion systems [52], and language-model-driven chemical embedding techniques [54] offer promising pathways to generalize the current approach toward broader chemical regimes, including transition state estimation and real-time reactive force field tuning [55].

**Acknowledgement**

The authors express their gratitude to Monolith AI for granting permission to utilize their Monolith platform. Mr. Amirali Shateri likes to acknowledge the University of Derby for the PhD studentship (contract no. S&E_Engineering_0722) and the support provided.